\documentclass{bmcart}

\usepackage[utf8]{inputenc} 

\usepackage{graphicx}
\usepackage{graphics}
\usepackage{subfigure}
\usepackage{tikz}
\usetikzlibrary{shapes,arrows,chains,positioning}
\usepackage{tikz-qtree}

\usepackage{xcolor}

\usepackage[fleqn]{amsmath}
\usepackage{amssymb}
\usepackage{amsfonts}
\usepackage{mathtools}
\usepackage{array}
\usepackage{siunitx}
\usepackage{algorithmic}
\usepackage{float}

\usepackage{lmodern}
\usepackage{textcomp}

\usepackage[square,comma,numbers,sort&compress]{natbib}

\usepackage[colorlinks,linkcolor=blue]{hyperref}

\startlocaldefs

\graphicspath{{fig_EJPDS/}} 
\definecolor{IgorGreen}{RGB}{76,153,0}
\definecolor{IgorBlue}{RGB}{0,0,204}
\definecolor{IgorGreen1}{RGB}{0,153,77}
\definecolor{IgorPurple}{RGB}{103,0,204}

\endlocaldefs

\begin{document}
	
	\begin{frontmatter}
		
		\begin{fmbox}
			\dochead{Research}
			
			
			\title{Modeling of Information Diffusion on Social Networks
				with Applications to WeChat}
			
			\author[
			addressref={aff1,aff2},                   
			email={liuliang@nudt.edu.cn}   
			]{\inits{L}\fnm{Liang} \snm{Liu}}
			\author[
			addressref={aff2},
			email={B.Qu@tudelft.nl}
			]{\inits{B}\fnm{Bo} \snm{Qu}}
			\author[
			addressref={aff1},                 
			email={nudtcb9372@gmail.com}
			]{\inits{B}\fnm{Bin} \snm{Chen}}
			\author[
			addressref={aff2},
			email={a.hanjalic@tudelft.nl}
			]{\inits{A}\fnm{Alan} \snm{Hanjalic}}
			\author[
			addressref={aff2},
			corref={aff2},
			email={H.Wang@tudelft.nl}
			]{\inits{H}\fnm{Huijuan} \snm{Wang}}
			
			
			\address[id=aff1]{
				\orgname{College of Information System and Management, National University of Defense Technology}, 
				\postcode{410073}                                
				\city{Changsha},                              
				\cny{China}                                    
			}
			\address[id=aff2]{%
				\orgname{Faculty of Electrical Engineering, Mathematics, and Computer Science,Delft University of Technology},
				\street{Mekelweg 4},
				\postcode{2628 CD}
				\city{Delft},
				\cny{The Netherlands}
			}
			
			
			\begin{artnotes}
			\end{artnotes}
			
		\end{fmbox}
		
		\begin{abstractbox}
			
			\begin{abstract} 
				
				Traces of user activities recorded in online social networks such as the creation, viewing and forwarding/sharing of information over time open new possibilities to quantitatively and systematically understand the information diffusion process on social networks. From an online social network like WeChat, we could collect a large number of information cascade trees, each of which tells the spreading trajectory of a message/information such as which user creates the information and which users view or forward the information shared by which neighbours. In this work, we propose two heterogeneous non-linear models, one for the topologies of the information cascade trees and the other for the stochastic process of information diffusion on a social network. Both models are validated by the WeChat data in reproducing and explaining key features of cascade trees.
				
				Specifically, we firstly apply the Random Recursive Tree (RRT) to model the cascade tree topologies, capturing key features, i.e. the average path length and degree variance of a cascade tree in relation to the number of nodes (size) of the tree. The RRT model with a single parameter $\theta$ describes the growth mechanism of a tree, where a node in the existing tree has a probability $d_i^{\theta}$ of being connected to a newly added node that depends on the degree $d_i$ of the existing node. The identified parameter $\theta$ quantifies the relative depth or broadness of the cascade trees, indicating that information propagates via a star-like broadcasting or viral-like hop by hop spreading. The RRT model explains the appearance of hubs, thus a possibly smaller average path length as the cascade size increases, as observed in WeChat. We further propose the stochastic Susceptible View Forward Removed (SVFR) model to depict the dynamic user behaviors including creating, viewing, forwarding and ignoring a message on a given social network. Beside the average path length and degree variance of the cascade trees in relation to their sizes, the SVFR model could further explain the power-law cascade size distribution in WeChat and unravel that a user with a large number of friends may actually have a smaller probability to read a message (s)he receives due to limited attention.

			\end{abstract}
			
			\begin{keyword}
				\kwd{Information diffusion}
				\kwd{stochastic model}
				\kwd{social networks}
				\kwd{WeChat}
				\kwd{random recursive tree}.
				
			\end{keyword}
			
		\end{abstractbox}
		%
		
	\end{frontmatter}
	
	\section{Introduction}
	
	The rapid development of the Internet, smart phones and information technology has facilitated the boost of online social networks, such as Facebook, Twitter, Flickr, Digg and Sina Weibo. Such online social networks allow message, content and information in general to spread faster and wider than ever (e.g. retweeting) \cite{kietzmann2011social,Guille2013,Obar2015,zhang2016dynamics}. Understanding the features and dynamics of information diffusion in social networks is crucial for businesses to promote products, but also for governments to predict and even regulate public opinion \cite{Hughes2009,Kaplan2010,Khondker2011}. 
	
	Most empirical work on aforementioned social networks has focused on basic statistical analysis of the features of the social networks, of the content popularity or of the content/information diffusion trajectories \cite{Kwak2010,Bakshy2011,Banos2013,Taxidou2014,Goel2015, Bakshy2012,Cha2009,Ghosh2011,Bao2013,Feng2015,Li2015,Wang2016,zhang2016structure}. For example, the information diffusion trajectory on a social network can generally be represented by a cascade (tree), where the root is the source node that creates the information and where the links represent the information transmitting paths between the users. The average path length of a cascade tree \footnote{The average path length is the average number of links in the shortest path between two nodes.}, also called structural virality, in relation to the size of the tree, may indicate how deep (viral like information propagation) or shallow (broadcasting type of information diffusion) a cascade tree is \cite{bounova2012overview, Goel2015}. Time series analysis of e.g. content popularity over time has been explored to distinguish between different types of contents \cite{crane2008robust,wu2007novelty}. Machine learning techniques have also been applied to, for example, predict content popularity based on the previous popularity and the features related to the contents and users that shared the content \cite{Richier2014Bio,Cheng2016Do}.
	
	Stochastic models, such as \emph{cellular automata} \cite{Goldenberg2001}, \emph{Threshold models} \cite{Granovetter1978,Liopinion2013,Boopinion}, \emph{Susceptible Infected Recovered (SIR)} \cite{Hethcote2000,Pastor-Satorras2001,Newman2002,Feng2015}, and \emph{Linear Influence} \cite{Yang2010} have been studied to understand how the dynamics of information diffusion such as the spreading rate and the social network topology could influence a key feature of the diffusion process such as the popularity. However, we still insufficiently understand to whether such first order models with few parameters could quantitatively reproduce several key features of real-world information diffusion.
	
	In this paper we aim to propose stochastic models with few parameters for (a) the topology of the information cascade trees and (b) the dynamics of information diffusion on a social network, that could capture several key features observed in the real-world cascade trees. Our models will be applied and validated by the WeChat dataset. WeChat is the most popular smart phone application in China and has about 800 million monthly active user accounts\cite{Tencent2016}. Apart from some sporadic efforts (e.g. \cite{Li2016}), information diffusion in WeChat has, not been studied extensively. Our choice of WeChat data to illustrate and validate our modelling methods is also because:
	\begin{itemize}
		\item WeChat is a semi-closed social network where information is shared mainly via strong social ties (i.e. friends that mutually agree to share information)
		\item WeChat records both the view and forward actions of each user and also from which node the information arriving to the user has been forwarded.
	\end{itemize}
	
	Topologies of cascade trees have so far been characterised by their average path length in relation to their sizes \cite{bounova2012overview, Goel2015}. The size of a cascade tree, i.e. the number of users that a message has reached in total in a social network, may range from hundreds to millions. Although a weak (strong) correlation between the average path length and the size of a tree may suggest shallow (deep) tree structures \cite{Anderson2015}, we lack a systematic method to quantify the shallowness or deepness of a group of cascade trees. In this work, we propose to use of the generalised random recursive tree (RRT) \cite{rudas2005random,krapivsky2001organization} with a single parameter to model cascade trees (possible of a given type of contents) with diverse sizes on an online platform. The RRT, a growth tree model, could well capture two features of WeChat cascade trees: the average path length and the degree variance, as a function of the cascade size. The identified parameter in the RRT model quantifies how deep or shallow the cascade trees are and implies the possible growing mechanisms of cascade trees. Via applying the RRT to the WeChat data, we find that hubs tend to appear as WeChat cascade trees become large in size, leading towards broadcast-like information diffusion. 
	
	Secondly, we propose the heterogeneous Susceptible View Forward Removed (SVFR) model, which allows users to have different probabilities of viewing a message, depending on their degree (the number of neighbours) in the underlying social network. Interestingly, our SVFR model could well explain the power-law distributed size of cascade trees, the degree variance and the average path length of a cascade tree in relation to the tree size as observed in the WeChat dataset. Importantly, our SVFR model benchmarked by the WeChat data points out that a WeChat user with a large number of friends tends to have a lower probability of viewing the messages shared by his/her friends, likely due to his/her limited energy and attention online. 
	
	The remainder of this paper is organized as follows: Section \ref{sec:data} describes the WeChat Moments diffusion data and how to construct the cascade trees. Section \ref{sec:rrt} and \ref{sec:svfr} present the RRT and SVFR model to capture the structure of the cascade trees and the dynamics of the information diffusion respectively. Section \ref{sec:conclusion} summaries our findings and points out interesting future work.
	
	\section{Dataset Description}
	\label{sec:data}
	
	We firstly introduce the information diffusion dataset from the WeChat\footnote{\url{https://en.wikipedia.org/wiki/WeChat}} platform\cite{Schiavenza2013,lien2014examining}, which will be used to validate the two models that we are going to propose. WeChat Moments (WM)\footnote{\url{https://en.wikipedia.org/wiki/Moments_(social_networking)}}, known as \emph{Friends Circle}, serves social networking functions in which users can view information shared by friends. In this work we focus on the diffusion of web pages in the WM network. A user may react to the web page forwarded/shared by his/her friend in three ways: (i) \emph{View} the web page, meaning that the user clicks the link of the web page and views the content (ii) ignore the web page without a click, and (iii) \emph{Forward} (or share) the URL of the web page to his/her friends.
	
	An example of the diffusion of a web page in the WM network is shown in Figure \ref{fig:tree}. Firstly, a user being at the root of the tree creates a web page and makes it available to his friends. Then his friends may ignore, view or forward the web page after seeing the web page appearing in their Friends Circle with a title. The forwarding of the information (web page) allows its friends to further view, forward or ignore the information.
	
	\begin{figure}[!htbp]
		\centering
		\begin{tikzpicture}
		[rootNode/.style={circle,draw=red!80,fill=red,very thick, minimum size=2mm},
		viewNode/.style={circle,draw=green!80,fill=green,very thick, minimum size=2mm},
		shareNode/.style={circle,draw=blue!80,fill=blue,very thick, minimum size=2mm},
		otherNode/.style={circle,draw=lightgray!80,fill=lightgray,very thick, minimum size=2mm},
		edge from parent/.style={black,draw,->},
		level 0/.style={sibling distance = 5mm, level distance = 15mm},
		level 1/.style={sibling distance = 5mm, level distance = 15mm},
		level 2/.style={sibling distance = 5mm, level distance = 15mm},
		level 3/.style={sibling distance = 5mm, level distance = 15mm},
		level 4/.style={sibling distance = 5mm, level distance = 15mm}]
		
		\node [rootNode](root){}	
		child foreach \x in {0,1,2,3} {node[viewNode]{}}
		child {node[shareNode]{}
			child foreach \x in {0,1,3,4} {node[viewNode]{}}
			child {node [shareNode] {}
				child {node [viewNode]{}}
				child {node [viewNode]{}}
				child {node [viewNode]{}}
				child foreach \x in {0,1,3,4} {node[otherNode]{} edge from parent[dashed]}}
			child foreach \x in {0,1,2,3,4} {node[otherNode]{} edge from parent[dashed]}
			child {node[shareNode]{}}
		}
		child foreach \x in {0,1,2,3,4,5,6,7,8} {node[otherNode]{} edge from parent[dashed]}
		child {node [shareNode] {}
			child foreach \x in {0,1,2} {node[viewNode]{}}
			child {node [shareNode] {}
				child {node [viewNode]{}}
				child {node [shareNode]{}
					child {node [shareNode]{}}
					child {node [shareNode]{}}
					child foreach \x in {0,1,2,3} {node[otherNode]{} edge from parent[dashed]}}
				child foreach \x in {0,1,2,3,4} {node[otherNode]{} edge from parent[dashed]}}
			child foreach \x in {0,1,2,3} {node[otherNode]{} edge from parent[dashed]}
		}
		child {node[shareNode](n1){}};
		
		\begin{scope}[xshift=2.5in,every tree node/.style={},edge from parent path={}]
		\Tree [.{depth 0} [.{depth 1} [.{depth 2} [.{depth 3} [.{depth 4} ]]]]]
		\end{scope}
		\end{tikzpicture}	
		\caption{\csentence{Schematic diagram of the diffusion of a web page in WeChat.} Colors differentiate between the users showing different behaviors regarding what they do with the information offered to them. The green circles represent users who have viewed the message. The blue circles stand for users who have shared the message after viewing it. The gray circles are those users who have not viewed the content. A (view) cascade tree is composed of the source node that creates the message, the nodes that have viewed the message, thus both the blue and green nodes and the black solid arrows among them. The (view) cascade tree of each web page is recorded in the data.}
		\label{fig:tree}
	\end{figure}
	
	We obtained the web page spreading dataset in WeChat Moments from a third-party service company\footnote{\url{http://www.fibodata.com/}}. The service company helps users create HTML5 format web pages to share the information including advertisements, web games, news, articles or holiday greetings. The dataset recorded from January $14$ to February $27$ in $2016$ all user activities, such as view and forward, and their corresponding time stamps related to all the web pages created with the format support from the service company. Both the content of the web pages and users are anonymised by web page indexes and user indexes, respectively. A user must first view a web page before (s)he forwards it. Whenever a user views a web page shared by a friend, the index of both the user who views the web page and the friend who shares the web page are recorded in the dataset, allowing us to construct the cascade tree for each web page. We select the web pages whose diffusion starts and ends within the period of 45 days. We assume that the diffusion of a web page stops if after the generation of the web page there is a day that no viewing/forwarding/sharing happens. Although resurgence of the spread of a content may occur after a silent day \cite{Anderson2015}, our assumption is supported by the data: given that a web page is neither forwarded nor viewed by any user in a given day, the probability that forwarding or viewing of this page occurs after that day within the $45$ days' observation window is $17\%$. We obtain $277,014$ web pages, whose life span is approximately within the considered time window. More than $7$ million users are involved in the diffusion of these web pages. For each web page, we construct its cascade tree, in which nodes represent the users who have viewed the web page and some of these nodes may have forwarded the web page. Such a cascade tree is also called a view cascade tree. Each information cascade is a tree without cycles because a user seldom views/forwards the same content more than once. If, in the rare case a user views (shares) a web page more than once, we consider only the first time when the user views (shares) the page. 
	
	These cascade trees collected from WeChat will be used to valid our models, i.e. whether our models could reproduce several key features of the observed cascade tree features.
	
	\section{Modeling of Information Cascade Tree Structure}
	\label{sec:rrt}
	
	In this section, we focus on the modelling of the topologies of the information cascade trees, without considering the underlying dynamics of users. We aim to propose a tree model that could construct trees that share similar properties of the cascade trees observed in WeChat. Firstly, we will analyse two fundamental properties of the information cascade trees in WeChat, that we would like our model to reproduce, namely the \emph{average path length} and \emph{degree variance}. Afterwards, we propose to use the Random Recursive Tree (RRT) model to model information cascade trees and illustrate to what extent this model could capture the two key features of the information cascades in WeChat.
	
	\subsection{Cascade Structure in WeChat}
	
	Two basic properties of a generic tree are the \emph{average path length} and the \emph{degree variance}. The average path length, also known as "Wiener Index" or "Structural Virality". It is defined as the average of the number of links $H_{ij}$ in the shortest path between any two nodes ${i}$ and ${j}$. Hence, in a tree with $N$ nodes we can formulate it as
	\begin{eqnarray}\label{wt}
	E[H] &=& \frac{1}{N(N-1)}\sum_{i=1}^{N}\sum_{j=1, j\neq i}^{N}H_{ij},
	\end{eqnarray}
	
	The \emph{degree variance} is the variance of degrees of all the nodes in a tree,
	\begin{eqnarray}\label{vt}
	Var[D] &=& \frac{\sum_{i=1}^{N}(d_i-E[D])^2}{N},
	\end{eqnarray}
	where the degree $d_i$ of the node $i$ tells how many links the node $i$ has and $E[D]$ is the average degree of all the nodes. The degree variance can be equivalently characterized by the standard deviation $\sqrt{Var[D]}$ of the degree, which is used later in our data analysis and model validation.
	
	It has been shown that the average path length is an important characteristic of information cascades and complex networks in general \cite{Wiener1947,Goel2015,Congmetrics}. Consider the class of cascade trees collected from an online social network. If the average path length of a cascade tree does not increase much with the size (number of nodes) of the tree, hubs may exist in relatively large cascade trees. In this case, information propagates via star-like broadcasting and large cascade trees are relatively shallow. However, if the cascade trees' average path lengths increase dramatically with their sizes, large cascade trees tend to be deep without large hubs and information spreads viral-like, hop by hop. 
	
	Both properties are sensitive to the size of the tree. For example, large trees may tend to have a large average path length. As shown in Figure \ref{fig:EmpAnat}, the sizes of the cascade trees collected from WeChat follow approximately a power-law distribution. Hence, we group the cascades trees according to their sizes that are slitted uniformly in logarithmic scale. We consider cascading trees that have more than $100$ nodes in the dataset, which corresponds to the web pages that could propagate to a certain extent. Both properties are explored for each group of trees. Figure \ref{fig:EmpBAT} (a) and (b) show the average path length and degree variance of a cascade tree (group) as a function of the size of the tree (group), respectively. The average path length increases as the size of the cascade tree increases, except that when the size is large. The decrease in the average path length around size $10^4$ is due to the hubs in the cascade trees, i.e. higher degree nodes, which is reflected in the large degree variance of large cascade trees.
	
	We aim to propose a tree model that could capture both properties as a function of tree size and could further quantitatively characterize how deep/shallow the cascade trees are.
	
	\subsection{The Random Recursive Tree Model}
	
	We propose to use the Random Recursive Trees RRTs to model the cascade trees. A RRT \cite{rudas2005random,krapivsky2001organization,kunegis2013preferential} is a growth tree model that starts with the root with node index $1$ at $t=0$ and adds a node $t+1$ at each time step $t$ to an existing node selected as follows: each existing node $i$ with its degree $d_i(t)$ at time t has the probability $ \frac{d_{i}^\theta(t)}{\sum_{i =1}^{t}d_{i}^\theta(t)}$ of being connected to the newly added node. Hence, the probability that an existing node is connected to a newly added node is proportional to the degree of this node of power $\theta\in[0,\infty)$. We denote a RRT with $N$ nodes and the scaling parameter $\theta$ by $T(N,\theta)$. Specifically, $T(N,0)$ corresponds to a uniform recursive tree (URT) where at each time step, a randomly selected existing node is connected to the newly added node \cite{su2006uniform,VanMieghem2014}. $T(N,1)$ is a scale-free tree where at each time step, the probability for an existing node to be connected to the new node is proportional to the degree of this node \cite{Szabo2002}. When $0<\theta<1$ ($\theta>1$), the probability that an existing node is attached to a new node is sub-linear (super-linear) of the degree of the existing node. When $\theta\to\infty$, the RRT approaches a star topology, whose average path length is $2-2/N$ for a star with $N$ nodes. 
	
\begin{figure}[!htb]
		\centering
		\subfigure[]{
			\includegraphics[scale=.5]{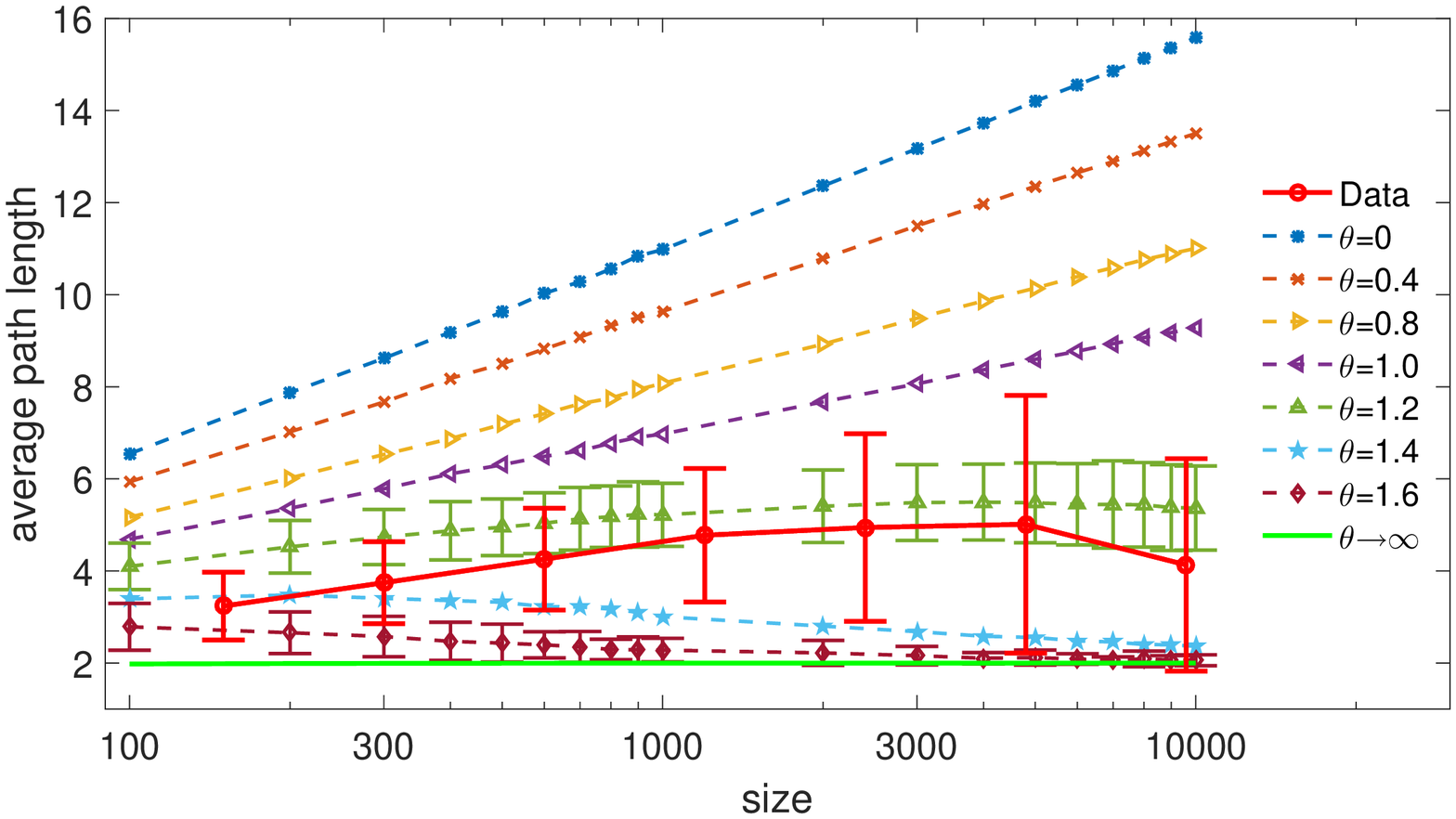}
			\label{fig:EmpBATa}
		}
		\subfigure[]{
			\includegraphics[scale=.5]{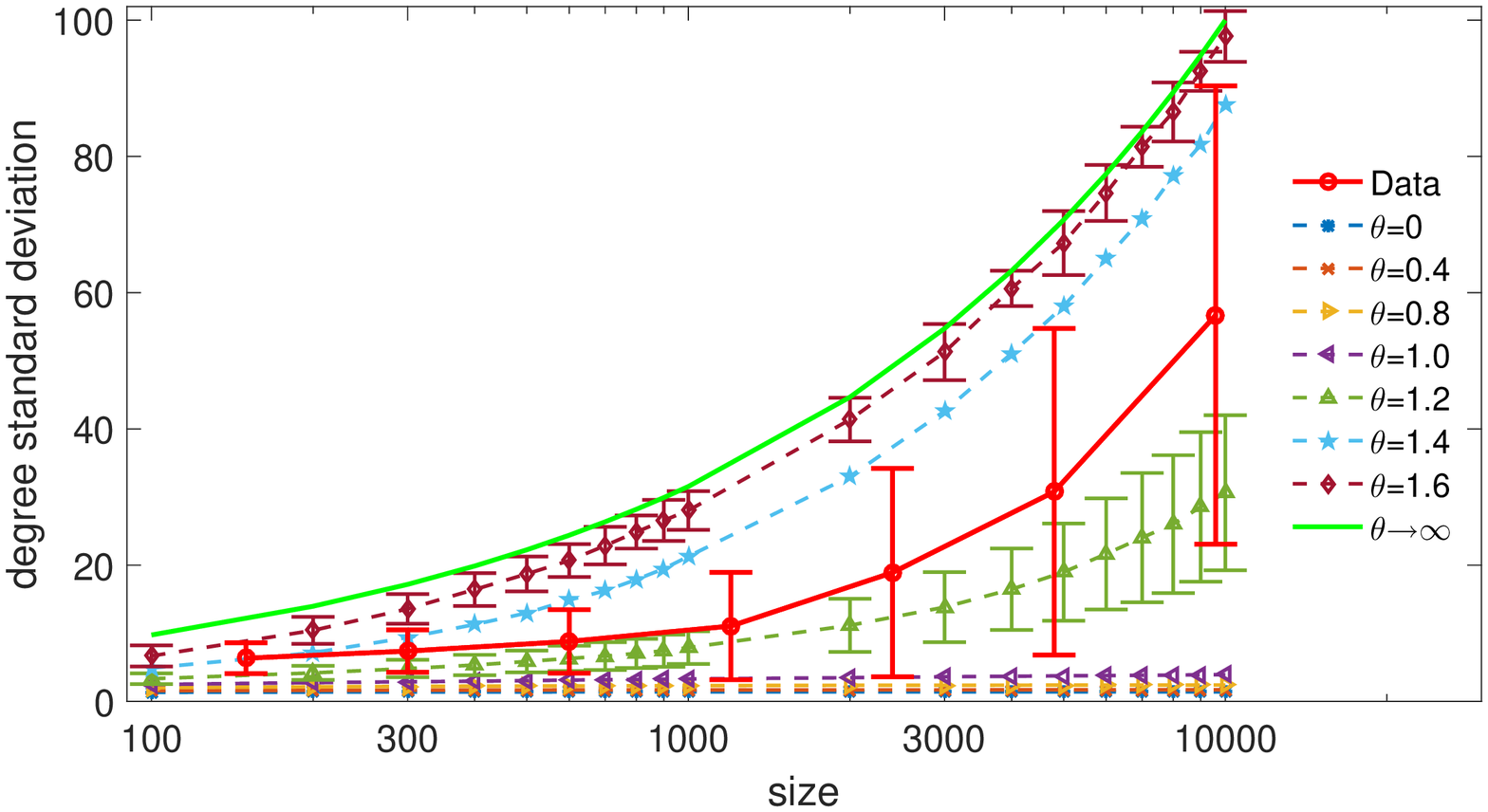}	
			\label{fig:EmpBATb}	
		}			
		\caption{\csentence{The average path length and degree standard deviation of the cascade trees in WeChat and the RRT models as a function of tree size.} The cascade trees in WeChat are grouped according to their sizes: [100,200), [200,400), [400, 800), [800,1600) etc. The average and standard deviation (error bar) of these two properties are obtained for each group and plotted as a function of the medium size of each group. For a given size of the trees and a given $\theta$, 1000 RRTs are generated independently and the average and standard deviation (error bar) of the average path length and degree standard deviation are obtained from the 1000 realization. The error bar for the two properties are shown for the RRT model with $\theta=1.2$ and $\theta=1.6$. }
		\label{fig:EmpBAT}
	\end{figure}

	We conduct $1000$ independent realisations of each RRT class $T(N, \theta)$ with size $N$ and scaling parameter $\theta$ and obtain for each class the average as well as the standard deviation of the two key topological features, i.e. the average path length and the degree variance. As illustrated in Figure \ref{fig:EmpBAT}, a small (large) $\theta$ suggests a relative deep (shallow like a star) tree with a large (small) average path length, that corresponds to the viral (broadcast) type of information diffusion.
	
	Figure \ref{fig:EmpBAT} shows that the average path length and degree variance, equivalently reflected by the degree standard deviation, in WeChat cascade trees as a function of the tree size can be well captured by RRT model with the scaling parameter $\theta$ around $1.2$ if we look at the mean of these two properties. When the variance of these properties, i.e. error bar, is taken into account, the WeChat cascade trees can be well described by the RRT model with $\theta>1$. This observation suggests that the WeChat cascade trees may follow a growth rule where a high degree node in the tree has a high probability to attract the connection to new nodes, such that large trees tend to have hubs, a large degree variance and a moderate average path length. When $\theta$ is positive, the average path length of RRTs increases first and decreases afterwards as the size of the tree increases. This can be observed evidently in the RRTs when $\theta=1.2$ in Figure \ref{fig:EmpBAT}. The average path length starts to decrease at a small tree size if $\theta$ is large. Such a change of the average path length as a function of the tree size is due to the fact that as a RRT grows with a positive $\theta$, hubs tend to appear and have a significantly higher chance to be connected to newly added nodes, and thus reduce the average path length and increase the degree variance. The average path length in WeChat cascade trees indeed increases first and then decreases as the cascade tree size increases, which can be thus well captured by the RRT model.
	
	The RRT model could be used to model the cascade trees, not limited to WeChat, that have diverse sizes. The parameter $\theta$ that best fits the data reflects quantitatively how deep the tree is and how diverse the degrees of the nodes in the tree are. In this way, we could compare different online systems with respect to in which system information propagates more via hubs/broadcasting or viral-like spreading.
	
	\section{Modeling of Information Cascade Process}
	\label{sec:svfr}

	In this section, we aim to develop a stochastic model of the information diffusion process. We develop this model based on our understanding of the WeChat information diffusion mechanisms and validate the model according to three key features observed in the WeChat dataset: the distribution of the sizes of the cascade trees, the average path length and the degree variance of a cascade tree in relation to the size of the tree.
	
	\subsection{The Susceptible View Forward Removed Model}
	We propose the Susceptible View Forward Removed (SVFR) model to describe the information diffusion process on a social network. This model is based on classic viral spreading models such as SIR model but more general and practical with respect to the definition of the possible states of a user and the possible non-liner and non-homogeneous probability for a user to view a message shared by its friend. 
	
	In the SVFR model, each node can be in one of the following four states at any time step: 
	\begin{itemize}
		\item Susceptible (S) - the user has the potential to read a message/content, but has not yet read it, 
		\item View (V) - the user views the message,
		\item Forward (F) - the user forwards the message, 
		\item Removed (R) - the user ignores the message either because (s)he does not want to read the message or has already viewed or forwarded the message.
	\end{itemize}
	For a given message, all the nodes are initially susceptible, except for the node that firstly publishes/shares this message thus is in state F at step $t=0$. The state transition diagram has been shown in Figure \ref{fig:flowchar}. For any node that is in state F at any time step $t$, each of its susceptible neighbours in the social network has a probability $\beta$ to view the message at step $t+1$. Moreover, each neighbor that views the message has a probability $\gamma$ to forward the message immediately after reading, and thus transits to state F at step $t+1$. In other words, each neighbor of a node in state F at time t, has a probability $\beta*(1-\gamma)$ of being in state V (view but not forward) and a probability $\beta\gamma$ of being in state F (read and forward) and probability $1-\beta$ of being in state R (ignore the message without reading the content) at time step $t+1$. For any node in state V or F at any given time, this node will be in state R at the next time step. The diffusion process of a message stops when all the nodes are either in state S or R, thus when the system reaches the stable state.
	
	\begin{figure}[!htbp]
		\centering
		\begin{tikzpicture}	[
		>=triangle 60,
		start chain=going right, 
		node distance=13mm, 
		every join/.style={norm}, 
		]
		\tikzset{
			base/.style={draw, on chain, on grid, align=center, minimum height=4ex},
			proc/.style={base, rectangle, text width=5em},
			test/.style={base, diamond, aspect=2, text width=5em},
			term/.style={proc, rounded corners},
			coord/.style={coordinate, on chain, on grid, node distance=6mm and 25mm},
			nmark/.style={draw, cyan, circle, font={\sffamily\bfseries}},
			norm/.style={->, draw, lcnorm},
			free/.style={->, draw, lcfree},
			cong/.style={->, draw, lccong},
			it/.style={font={\small\itshape}}
		}
		\node [proc, fill=green!25] (S) {Susceptible};
		\node [proc, fill=green!25] (V) {View};
		\node [proc, fill=green!25] (F) {Forward};
		\node [proc, fill=green!25] (R) {Removed};
		\node [coord, above = of S] (sa) {};
		\node [coord, below = of V] (vb) {};
		\node [coord, above = of R] (ra) {};
		\node [coord, below = of R] (rb) {};
		\path (S.east) to node [midway,above] {$\beta$} (V);
		\draw [->,black] (S.east) -- (V);
		\path (V.east) to node [midway,above] {$\gamma$} (F);
		\draw [->,black] (V.east) -- (F);
		\path (F.east) to node [midway,above] {$1$} (R);
		\draw [->,black] (F.east) -- (R);
		\path (S.north) to node [midway,yshift=1.8em] {$1-\beta$} (R.north);
		\draw [->,black] (S.north) -- (sa) |- (ra) -| (R.north);
		\path (V.south) to node [midway,yshift=-1.8em] {$1-\gamma$} (R);
		\draw [->,black] (V.south) -- (vb) |- (rb) |- (R.south);
		\end{tikzpicture}	
		\caption{\csentence{States transition diagram of the SVFR model.} }
		\label{fig:flowchar}
	\end{figure}
	Furthermore, we generalize the SVFR model to be a heterogeneous stochastic model where the probability $\beta$ that a user reads a message shared by its friend may depend on the degree of this user in the underlying social network. This is motivated by the fact that a node has a large number of friends tends to have a low probability to read a message shared by his/her friend due to the large number of messages he/she is exposed to and his/her limited effort in reading messages \cite{DUNBAR,Perra2011}. Without loosing generality, we assume that the probability for a node $i$ to read a message shared by a neighbor $\beta_i = cd_{i}^{-\alpha}$ may depend on the degree $d_{i}$ of this node, where the power exponent $\alpha$ is assumed to be positive and the constant $c$ is determined by the given average probability $\beta$ to view a message over all the nodes \footnote{Each node may view a message maximally once.}:
	\begin{eqnarray}\label{viewPro}
	\beta &=& c\sum_{k=d_{min}}^{d_{max}}k^{-\alpha} Pr[D=k],
	\end{eqnarray}
	
	As observed in the data and assumed in our model, users seldom reads or share a message more than once. The average view probability $\beta$ suggests how infectious/interesting a message is for users to view it. When $\alpha=0$, all nodes have the same view probability. Similar homogeneity has been usually assumed in previously proposed information diffusion models \cite{Goel2015}. Our heterogeneous model takes into account the possibility that the view probability of each node may be inversely proportional to the degree of the node, characterised by the degree scaling parameter $\alpha$. In the proposed stochastic model, we did not take into account a realistic and possibly heterogeneous time delay, e.g., between the time when a node shares a message and the time a neighbor reads or shares the message. 
	
	We assume that the probability $\gamma$ that a user forwards a message after viewing it, the so-called forward probability, is a constant, which is a simple start for the model study. Given the underlying social network and given the parameters $\alpha$, $\gamma$ and $\beta$ to be calibrated, the SVFR model could iterate the stochastic propagation of a message, each resulting in a cascade tree composed of users that have created, viewed and forwarded the message.

	\subsection{Model Validation}
	\begin{figure}[!htb]
		\centering
		{
			\includegraphics[scale=.8]{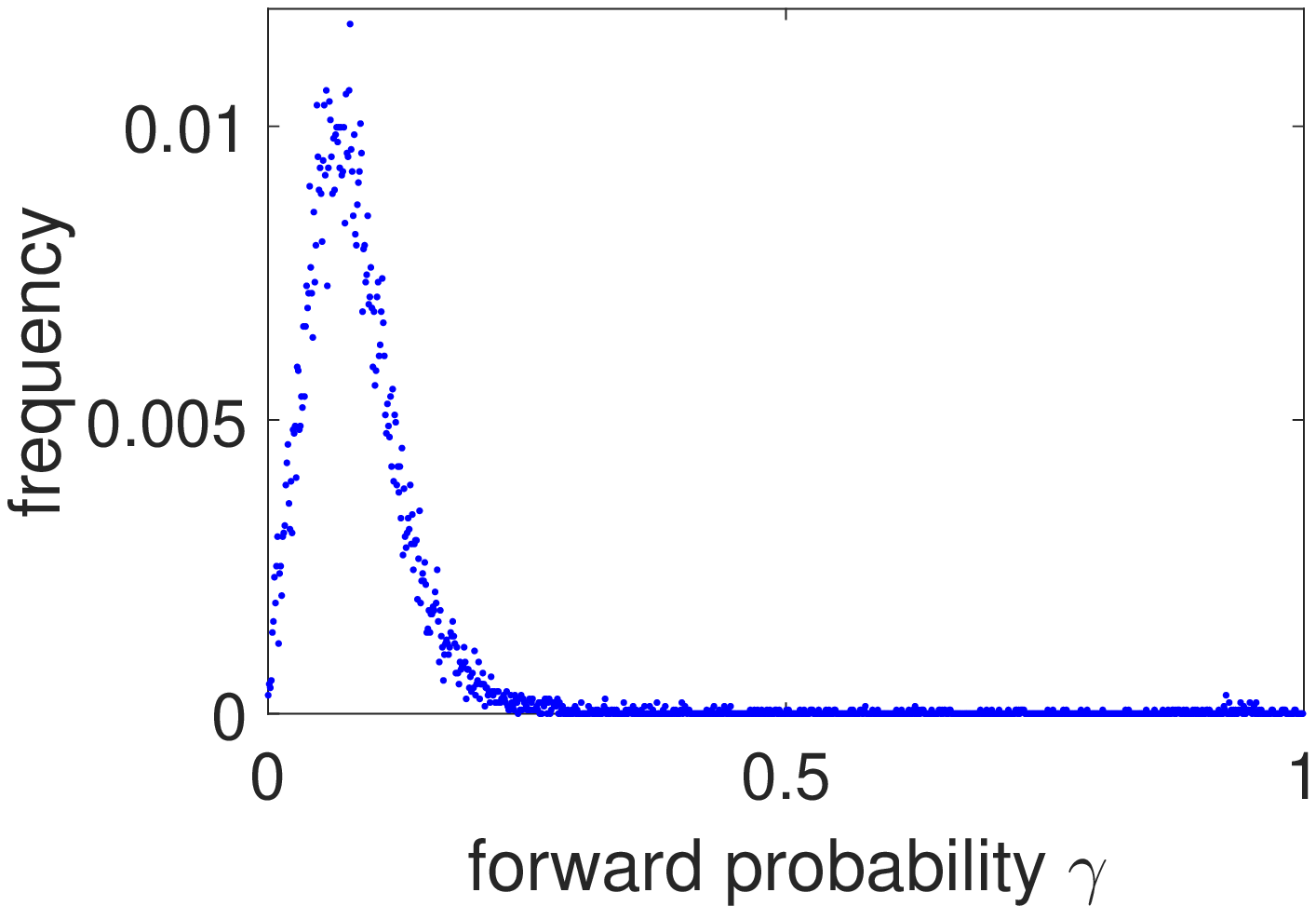}	
			\label{fig:EmpAna8a}	
		}
		\caption{\csentence{Distribution of the average forward probability in a cascade tree.} This distribution is obtained from the WeChat cascade trees that have a size larger than or equal to $100$.}
		\label{fig:EmpAna}	
	\end{figure} 
	The (average) forward probability in a cascade tree can be obtained as the number of nodes that forward the message over the total number of nodes in the cascade. Figure \ref{fig:EmpAna} shows that the forward probabilities of the WeChat cascade trees follows approximately a Gaussian distribution where forward probabilities are close to the average. Hence, we consider the average forward probability $\gamma=0.091$ observed in the data as the forward probability in our SVFR model.	
	
	The WeChat social network topology is unknown. Hence, we cannot derive directly from the data the two parameters related to the degree dependent view probabilities: the average view probability $\beta$ and scaling parameter $\alpha$. Instead, we will explore whether the SVFR model with tunable parameters $\beta$ and $\alpha$ could reproduce the three key features of the WeChat cascade trees: the size distribution, the average path length and degree variance in relation to the tree size. The distribution of the sizes of the cascade trees is a crucial feature for a online social network, characterizing the distribution of the prevalence or popularity of the information propagated on the network. 
	We assume that the underlying social network is a scale-free network with a power law degree distribution 
	$Pr[D = k] = ck^{-\phi}$, as observed in many real-world networks \cite{Barabasi1999}. We use the configuration model \cite{Newman2005,Clauset2009,Hernandez07aqualitative} to construct the random scale-free networks with a power exponent of the degree distribution $\phi=2.5$, a minimum degree $d_{min}=10$ as in \cite{Goel2015} and a cutoff of the maximum degree $d_{max}=N^{1/(\phi-1)}$ \cite{Cohen2000}, where $N$ is the network size. When the network size is $N=10^5$, the average degree $E[D]\approx26.7$.

	For each given pair of $\beta$ and $\alpha$, we generate independently $100$ scale-free networks using the configuration model with $N=10^5$ nodes and power exponent $\phi=2.5$. On each generated network, we carry out the information spread of 100 messages independently according to the SVFR model where the initial node that creates/shares the message is chosen uniformly at random. In total, we obtain $10^{4}$ cascade trees for the given $\beta$ and $\alpha$.
	
\begin{figure}[!htb]
		\centering
		{
			\includegraphics[scale=.8]{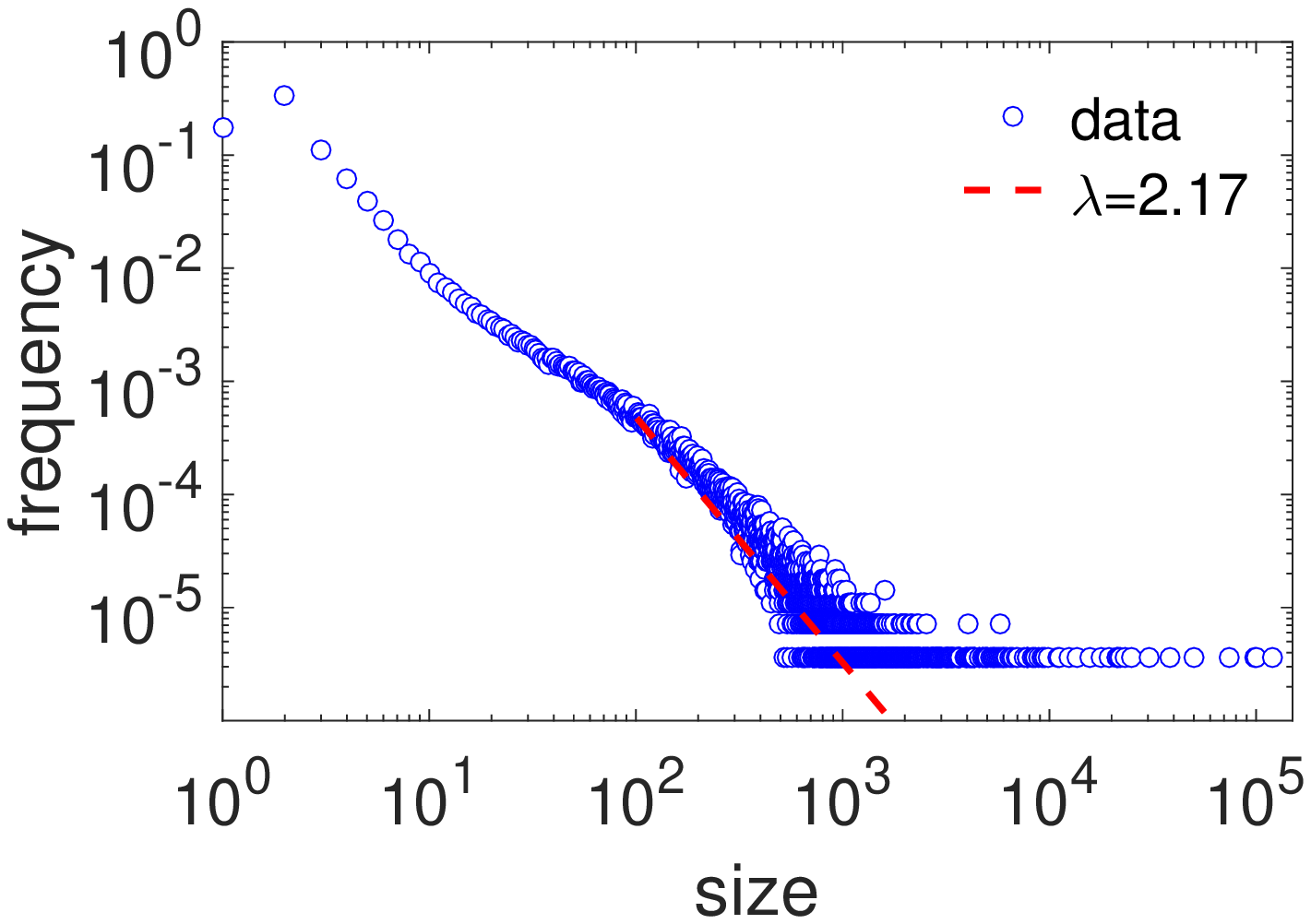}	
			\label{fig:EmpAna1at}	
		}
		\caption{\csentence{Distribution of the size of the WeChat cascading trees with the curve fitting for the tail where the size is larger than or equal to 100.}}
		\label{fig:EmpAnat}	
	\end{figure}

	Firstly, we explore the distribution of the sizes of the cascade trees in both the WeChat dataset and in our SVFR model. As shown in Figure \ref{fig:EmpAnat}, the distribution of the sizes of the observed WeChat cascade trees is approximately a power-law distribution. Since we are interested in the cascade trees with a size larger than $100$, that corresponds to the messages that could propagate to a certain extend, we fit the tail part of the distribution when the size is larger than or equal to $100$. The power exponent is approximately $\lambda = 2.17$. The power-law cascade size distribution has also been observed in other social networks, such as Twitter \cite{Taxidou2014,Banos2013,Goel2012}, Flickr \cite{Cha2009}, Digg \cite{Ghosh2011} and Sina Weibo\cite{Bao2013}. 
	
\begin{figure}[!ht]
		\centering
		\subfigure[]{
			\includegraphics[scale=.25]{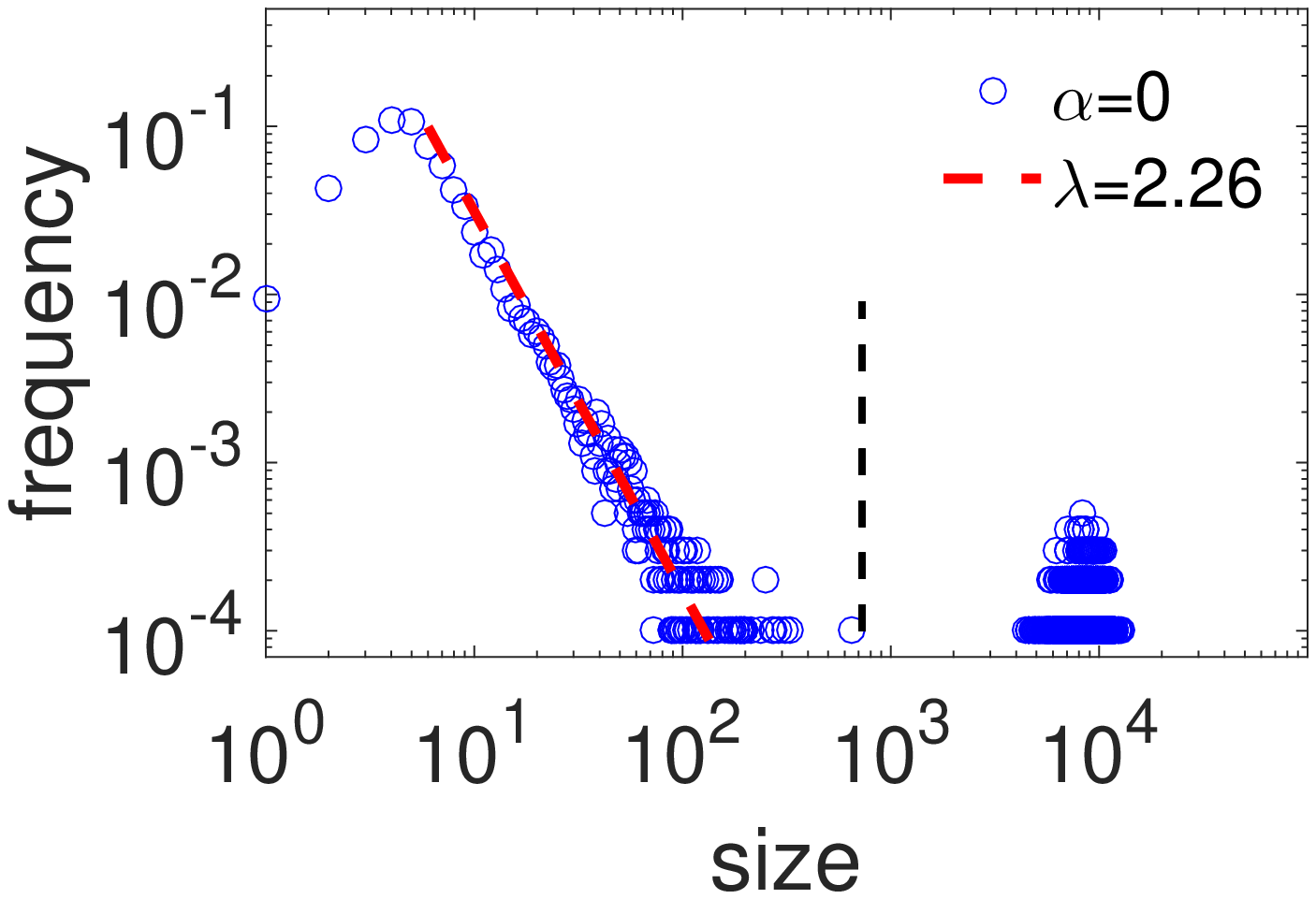}
			\label{fig:SeirSizeAna1a}		
		}
		\subfigure[]{
			\includegraphics[scale=.25]{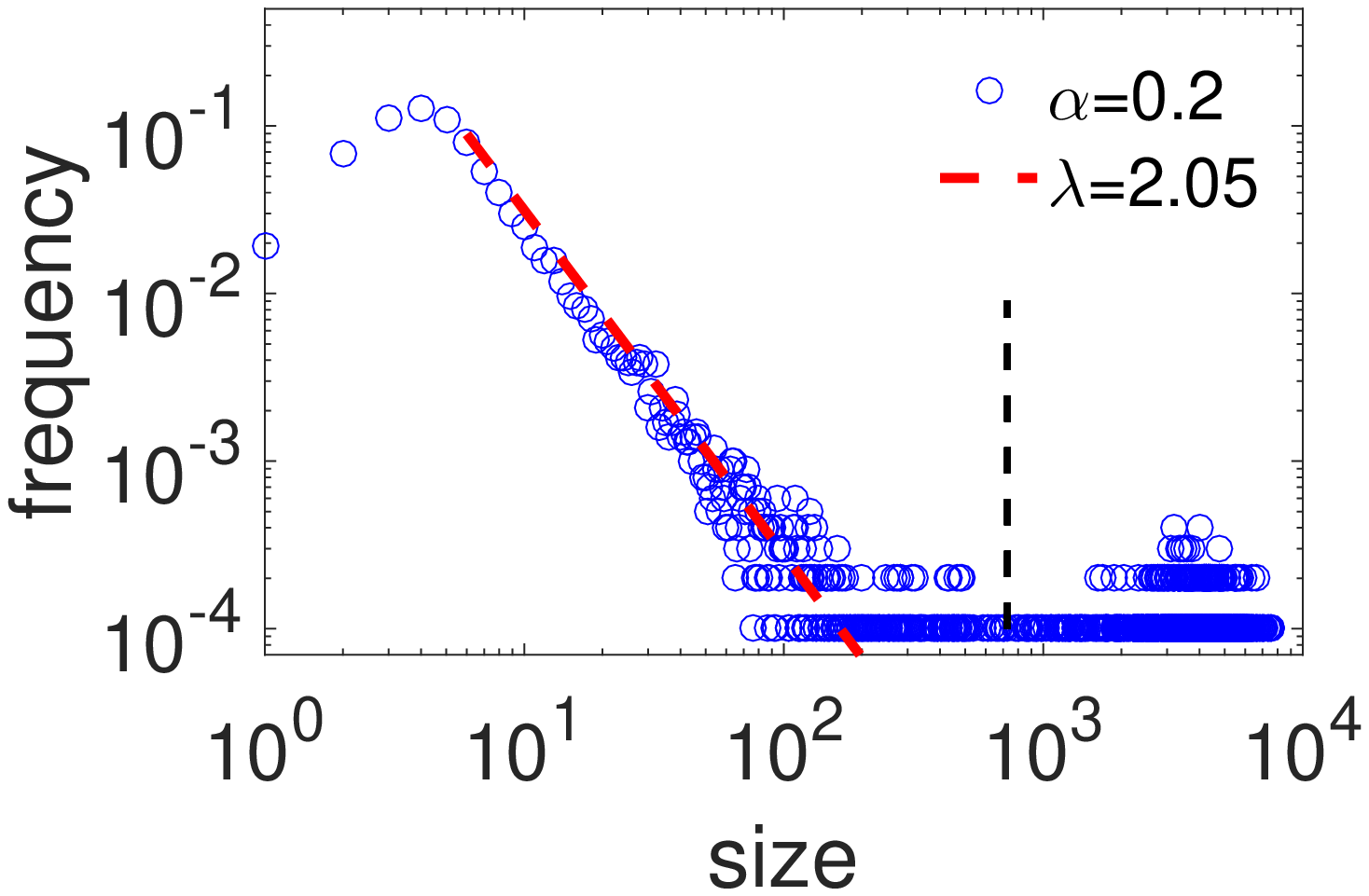}
			\label{fig:SeirSizeAna1b}
		}
		\subfigure[]{
			\includegraphics[scale=.25]{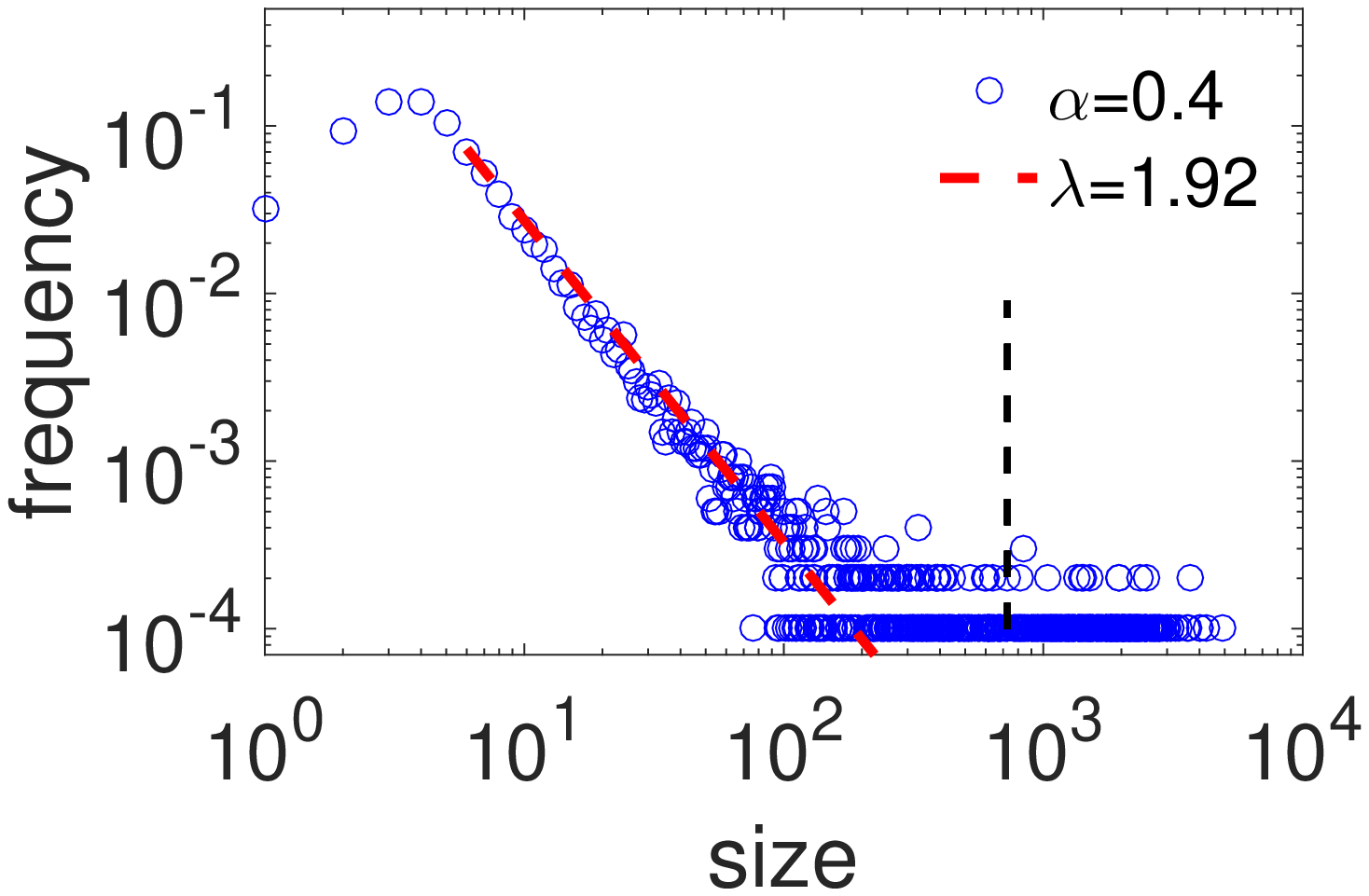}	
			\label{fig:SeirSizeAna1c}	
		}
		\subfigure[]{
			\includegraphics[scale=.25]{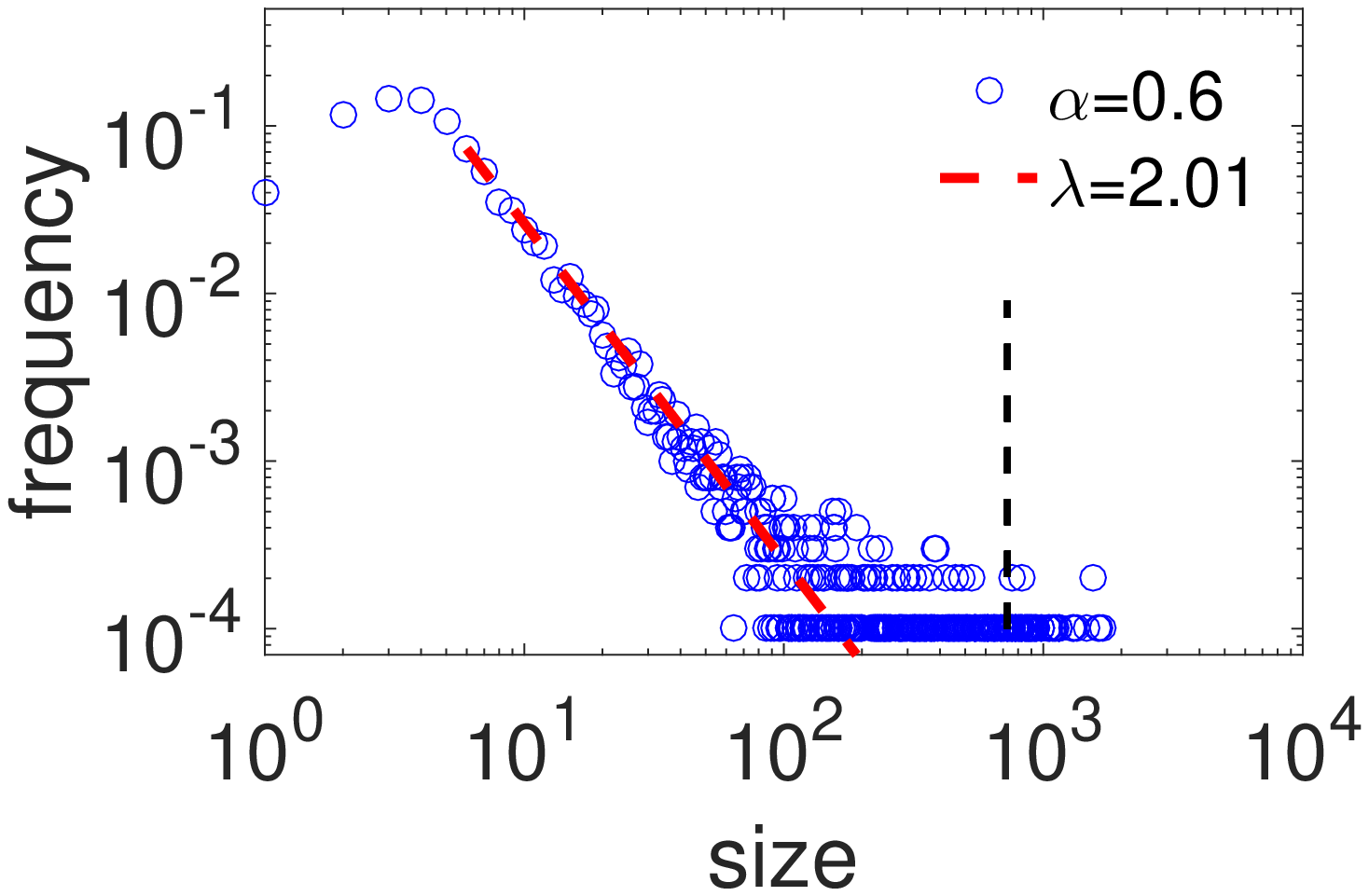}
			\label{fig:SeirSizeAna1d}
		}
		\subfigure[]{
			\includegraphics[scale=.25]{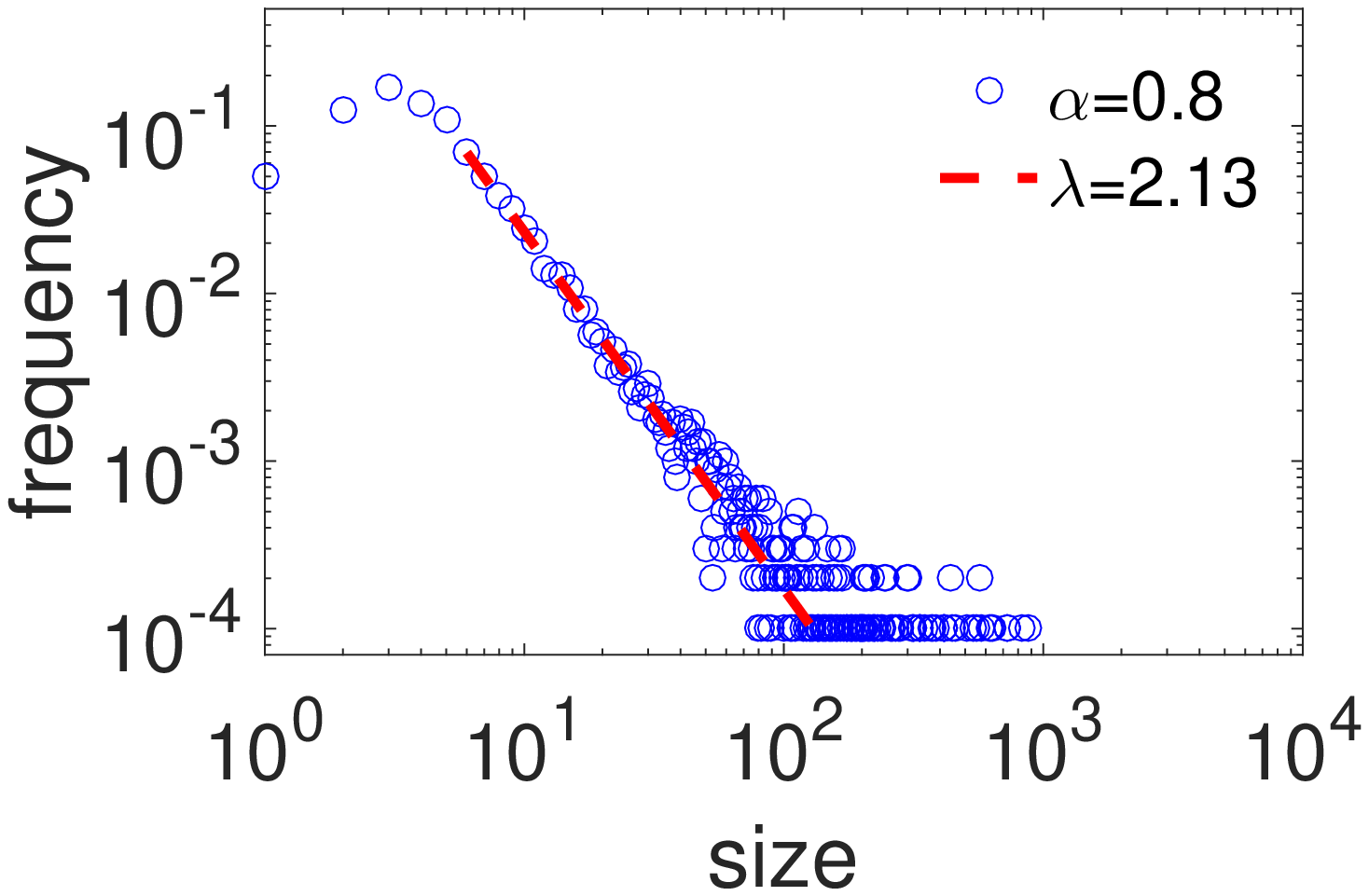}	
			\label{fig:SeirSizeAna1e}	
		}
		\subfigure[]{
			\includegraphics[scale=.25]{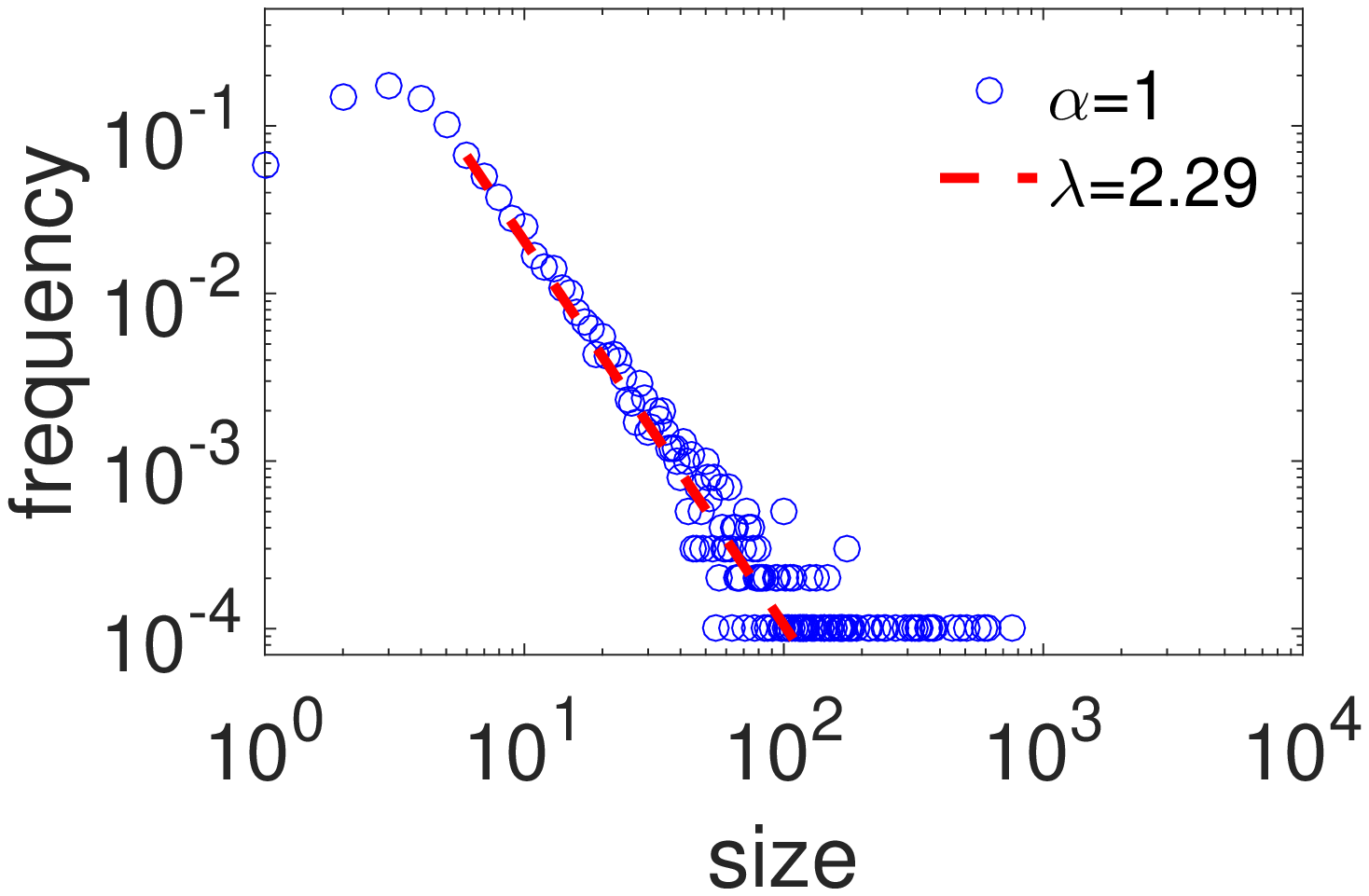}
			\label{fig:SeirSizeAna1f}
		}
		\subfigure[]{
			\includegraphics[scale=.25]{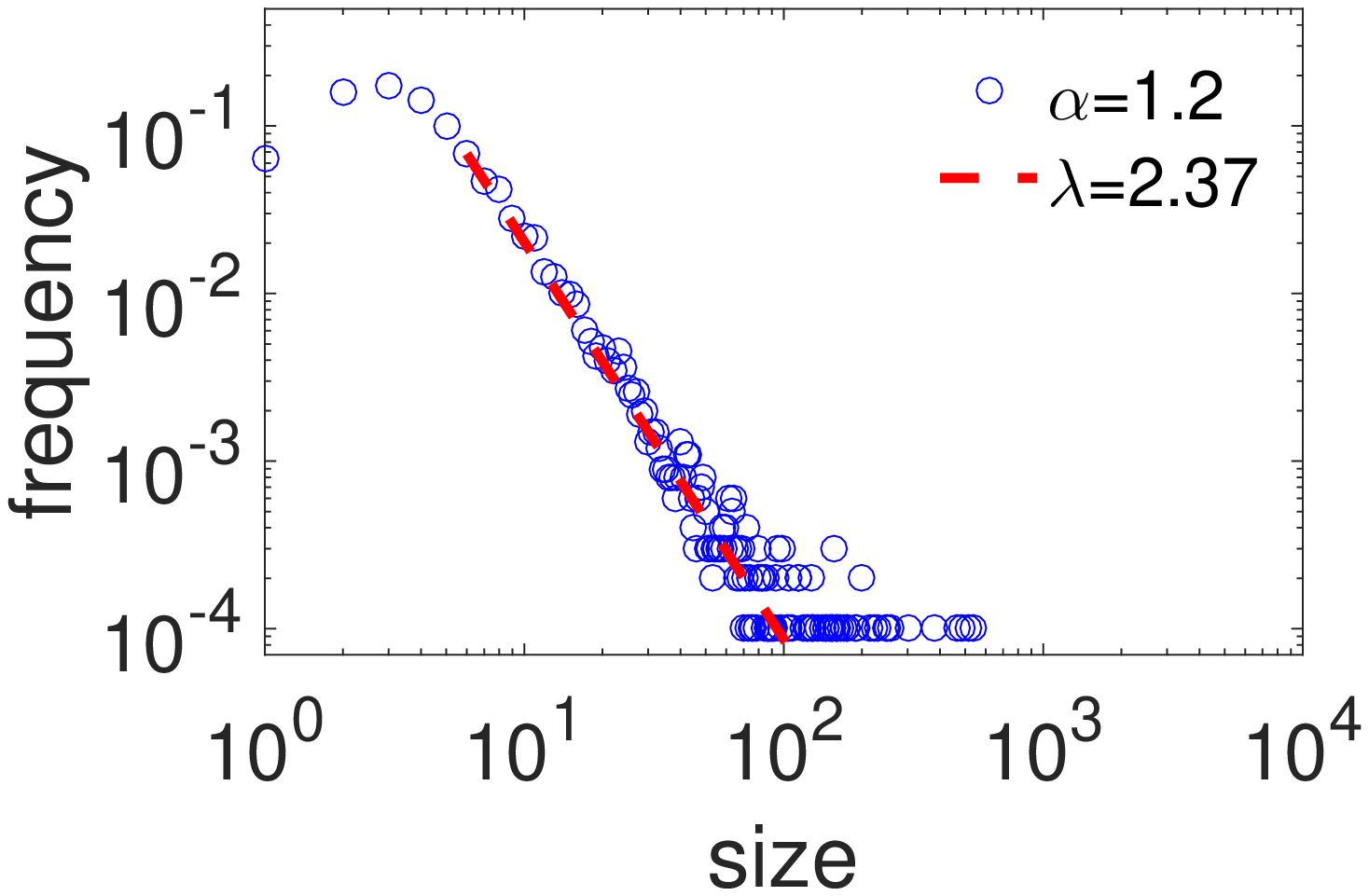}	
			\label{fig:SeirSizeAna1g}	
		}
		\subfigure[]{
			\includegraphics[scale=.25]{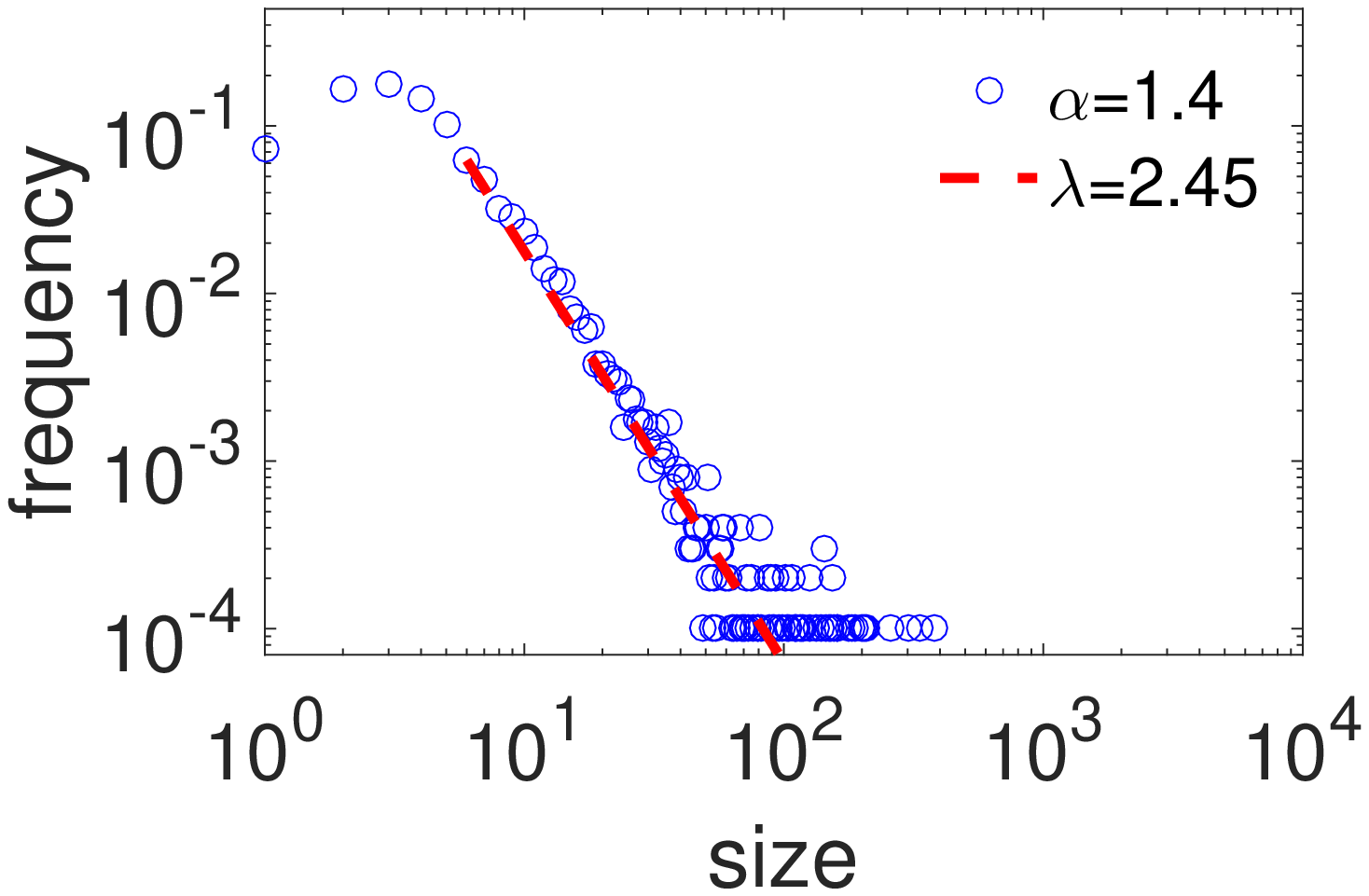}
			\label{fig:SeirSizeAna1h}
		}
		\subfigure[]{
			\includegraphics[scale=.25]{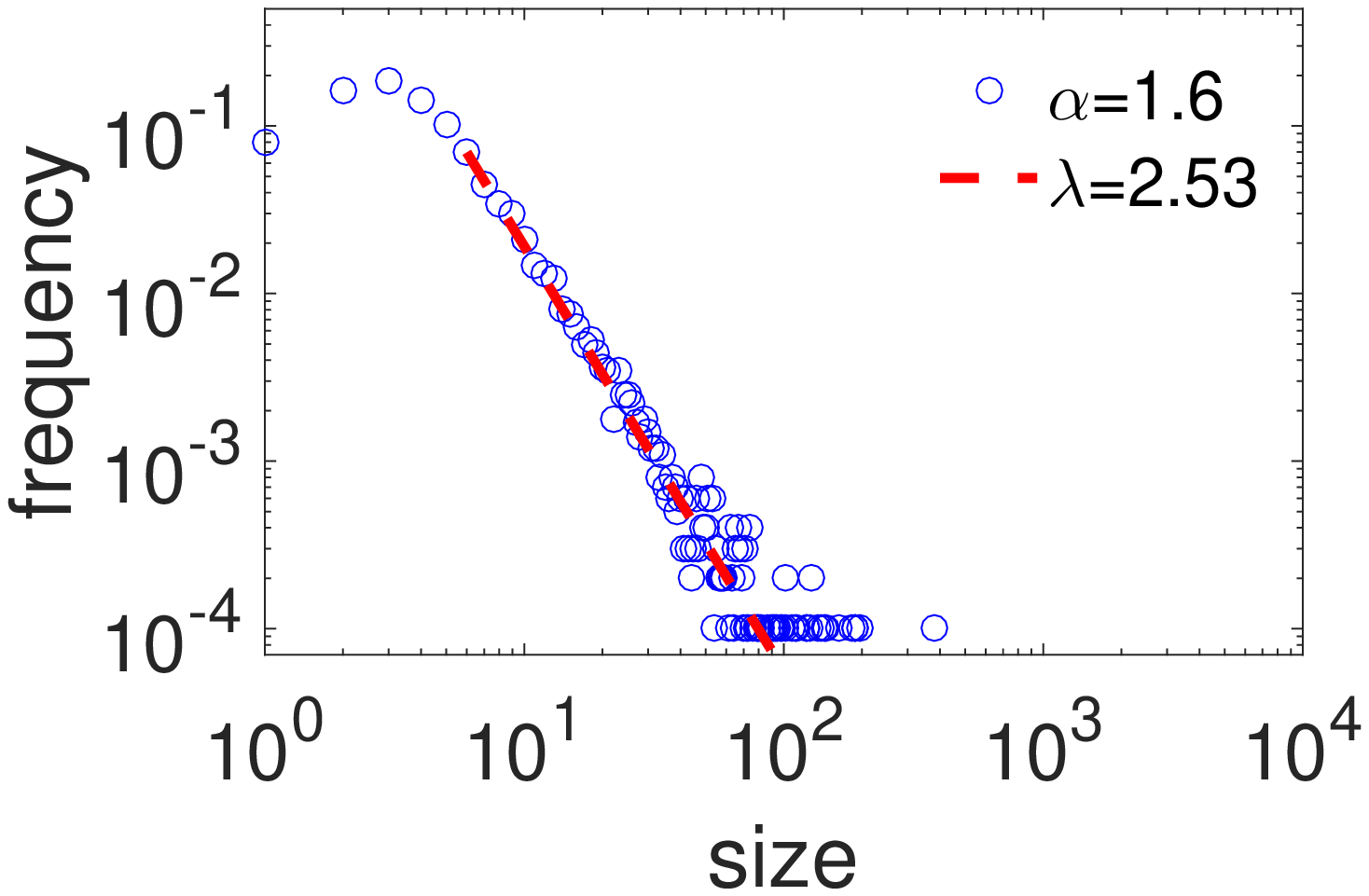}
			\label{fig:SeirSizeAna1i}
		}	
		\caption{\csentence{Cascade size distribution of the SVFR model for different degree scaling parameter $\alpha$.} The underlying scale-free network size is $N=10^5$ and the average view probability is $\beta=0.3$. The power-law part of the tail has been fitted. Each figure is obtained by $100$ independent realisations of the SVFR process on each of the $100$ independently generated underlying scale-free networks.}
		\label{fig:SeirSizeAna1}
	\end{figure}

	We take as an example the SVFR model with the average view probability $\beta=0.3$ whereas the degree scaling parameter $\alpha$ varies. Figure \ref{fig:SeirSizeAna1} illustrates how the size distribution of the cascade trees generated by our SVFR model changes as the degree scaling parameter $\alpha$ increases. 
	
\begin{figure}[!ht]
		\centering
		\subfigure[]{
			\includegraphics[scale=.4]{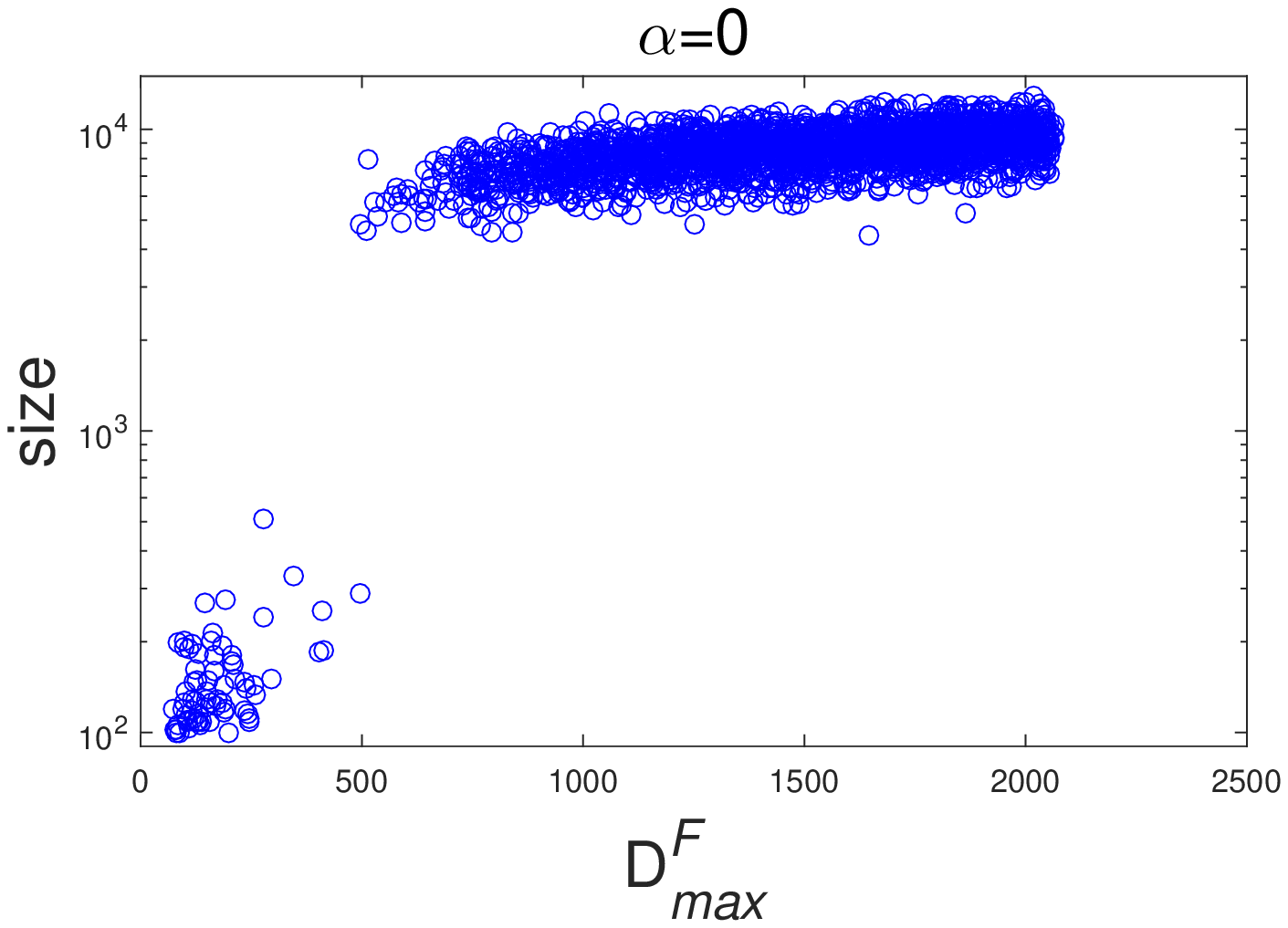}
			\label{fig:dega2a}		
		}
		\subfigure[]{
			\includegraphics[scale=.4]{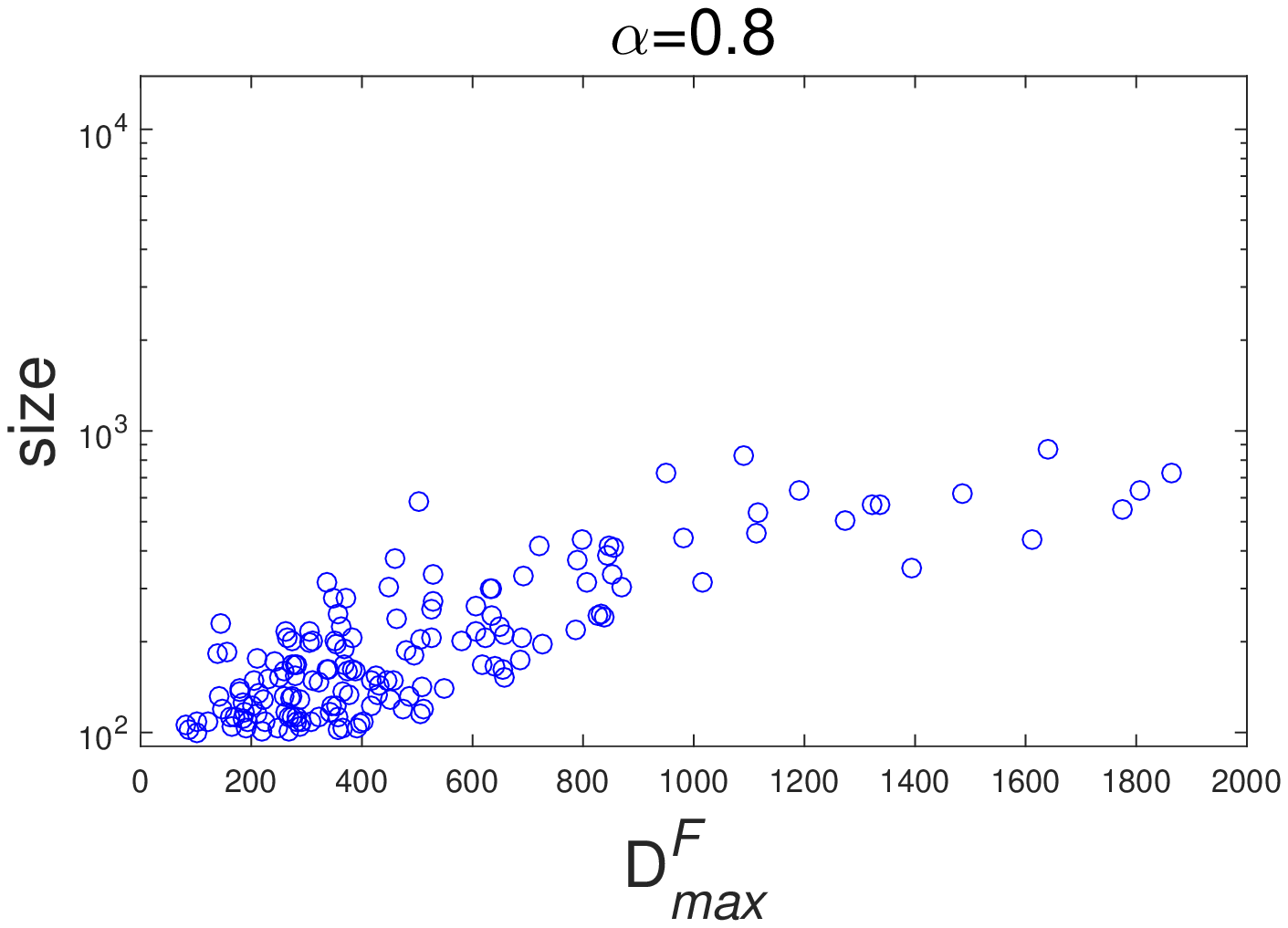}
			\label{fig:dega2b}
		}		
		\caption{\csentence{The size of a cascade tree generated by the SVFR model versus the maximum degree $D_{max}^F$ in the underlying social network of the nodes that have forwarded the message in the cascade tree when (a) $\alpha=0$ and (b) $\alpha=0.8$. Cascade trees larger than $100$ in size are considered.}}
		\label{fig:dmax}
	\end{figure}

	It turns out that when $\alpha=0$, i.e. when all the nodes have the same probability to view a message, the cascade size distribution has a peak in the tail. In this case, the cascade size of our model does not follow a power-law distribution as WeChat cascades but has a significantly higher probability to be large. Similar observation holds when the degree scaling parameter $\alpha$ is small. When the view probability $\beta$ or the network size $N$ increases, the separation between the power law decrease and the peak in the size distribution becomes even more apparent. As $\alpha$ increases, the cascade size distribution becomes a power-law distribution, the same as observed in WeChat. The hubs play a key role in such a change in the size distribution. Firstly, a hub (a high degree node in the underlying scale-free network) has a higher probability that one of its neighbors forwards the message than low degree nodes. Secondly, a hub has a higher probability to view thus forward a message when $\alpha$ is smaller and given the same average view probability $\beta$. Thirdly, the forwarding of a message by a hub allow its large number of neighbours to further view and forward the message, leading potentially to a large cascade. Hence, hubs facilitate the appearance of large cascades, especially when $\alpha$ is small. This explains as well why the largest possible cascade size decreases as $\alpha$ increases. Figure \ref{fig:dmax} further supports our explanation. We look into the maximal degree (in the underlying social network) $D_{max}^F$ of that nodes that have forwarded the information in a cascade tree in relation to the size of the cascade. As the $D_{max}^F$ increases, i.e. a higher degree node involves in the forwarding of the message, an abrupt jump occurs in the cascade size, when $\alpha=0$. Hence, the bulk in the size distribution $\alpha=0$ corresponds to the large cascades where hubs involve in forwarding the information.
	When $\alpha=0.8$, the increase of the cascade size with $D_{max}^F$ is relatively continuous. 
	Figure \ref{fig:SeirSizeAna1} suggests that $\alpha$ should not be small in order to capture the power-law size distribution in the WeChat dataset.

\begin{figure}[!h]
		\centering
		\includegraphics[scale=.65]{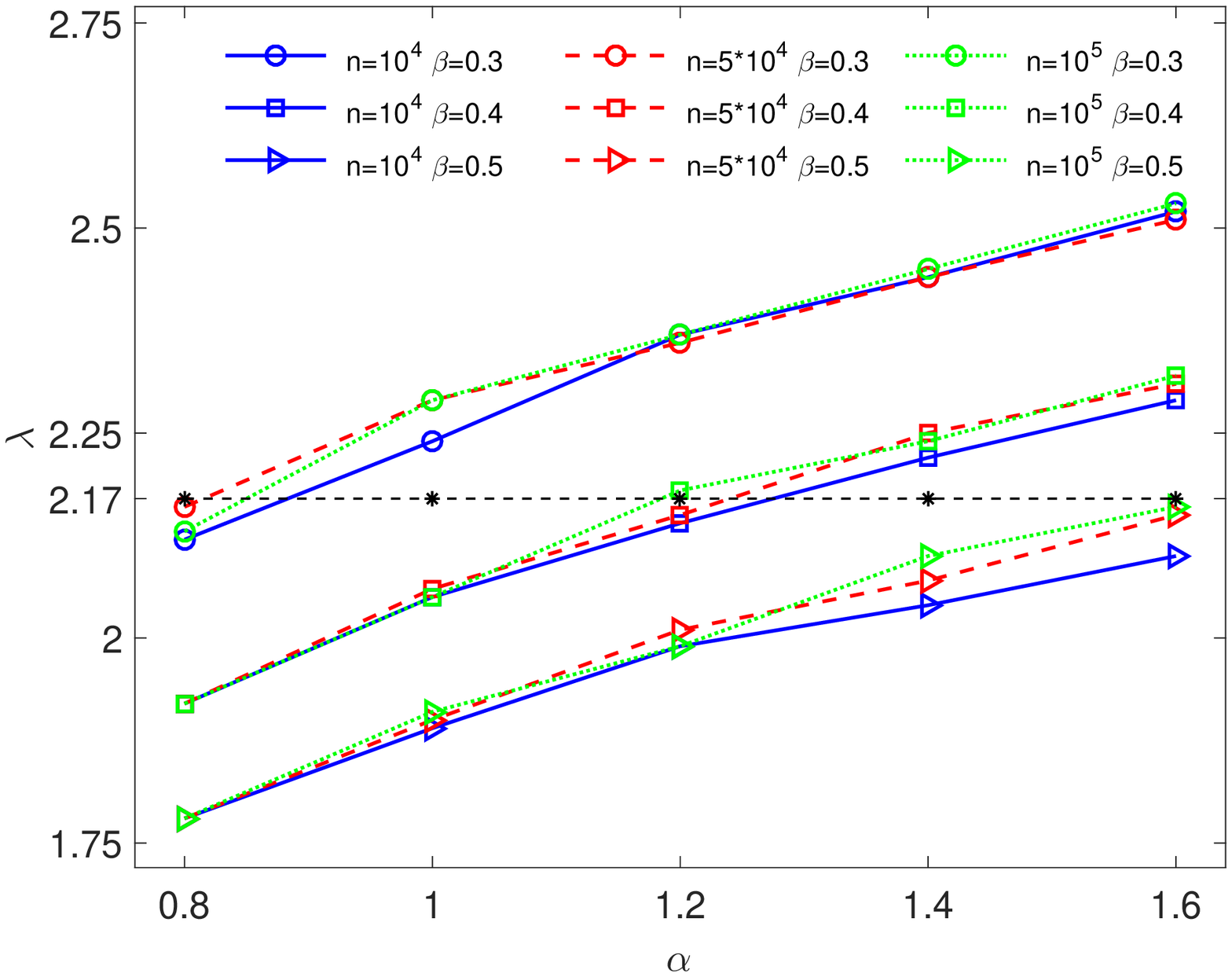}			
		\caption{\csentence{The power exponent $\lambda$ of the power-law cascade size distribution generated by the SVFR model as a function of the size $N$ of the underlying network, the average view probability $\beta$ and the degree scaling parameter $\alpha$.} For each set of parameters, the cascade size distribution is obtained from the 100 iterations of the SVFR information spread on each of the 100 independently generated underlying social networks.}
		\label{fig:SeirSizeAna2}
	\end{figure}

	Furthermore, we explore how the power exponent/slope $\lambda$ of the power-law cascade size distribution generated by the SVFR model is influenced by the size $N$ of the underlying network, the average view probability $\beta$ and the degree scaling parameter $\alpha$. As shown in Figure \ref{fig:SeirSizeAna2}, the exponent $\lambda$ is obtained via the power-law curve fitting of the power-law decreasing part of the size distribution. Although different curving fitting methods may influence the obtained power exponents \cite{Clauset2009}, we adopt this simplest method to illustrate our methodologies to identity the parameters of the proposed SVFR model.
	
	As shown in Figure \ref{fig:SeirSizeAna2}, power exponent $\lambda$ is insensitive to the size $N$ of the underlying networks, though the average cascade size may depend on the size of the underlying network. We will focus on the underlying network size $N=10^5$, which is large as well feasible for simulations. A smaller $\alpha$ and a large average view probability $\beta$ contribute to a smaller power exponent $\lambda$, thus large cascade trees with a higher probability. The power exponent $\lambda=2.17$ observed in WeChat can be approximated by our SVFR model when $\beta=0.3$ and $\alpha=0.8$ or $\beta=0.4$ and $\alpha=1.2$ or $\beta=0.5$ and $\alpha=1.6$.
	
	Finally, we investigate the average path length and the degree variance of the cascade trees in relation to the cascade tree sizes produced by our SVFR model with the aforementioned three sets of parameters that could already well capture the cascade size distribution of WeChat. 
	
	\begin{figure}[!h]
		\centering
		\subfigure[]{
			\includegraphics[scale=.4]{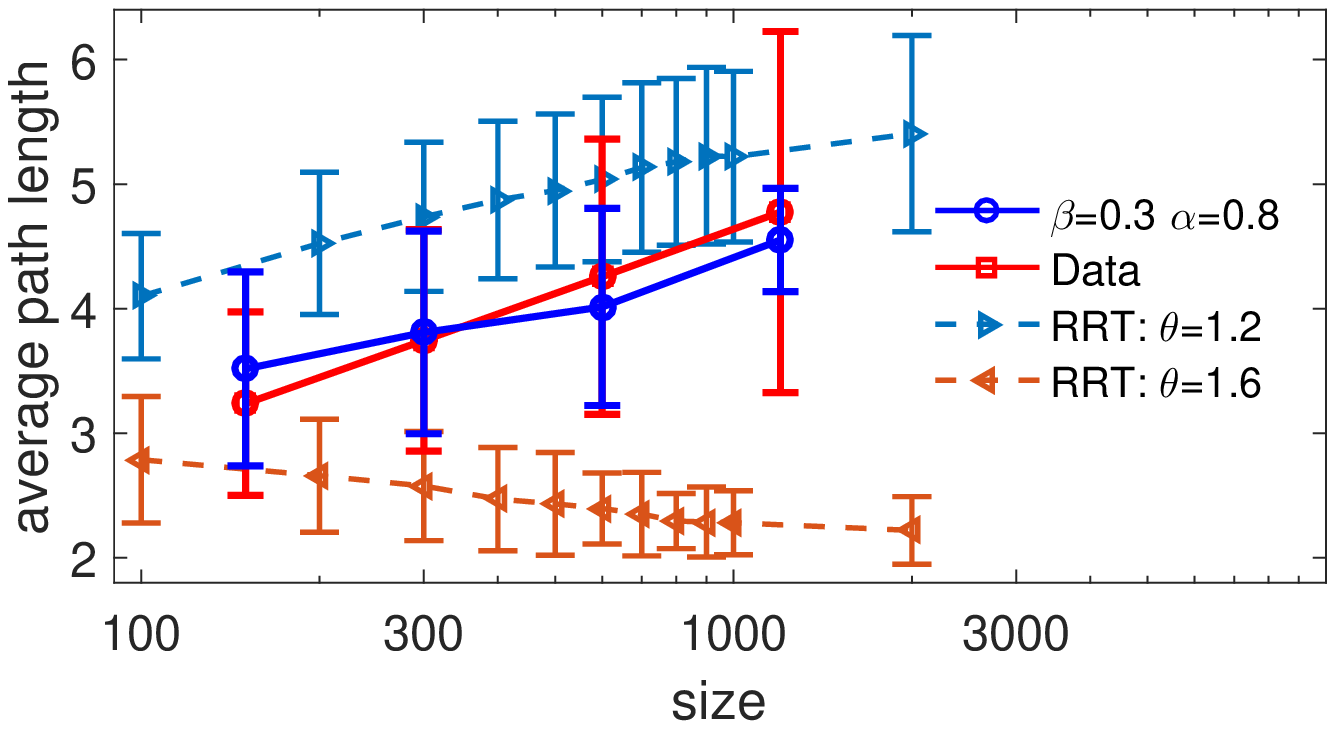}
			\label{fig:SeirTopoAna1a}		
		}
		\subfigure[]{
			\includegraphics[scale=.4]{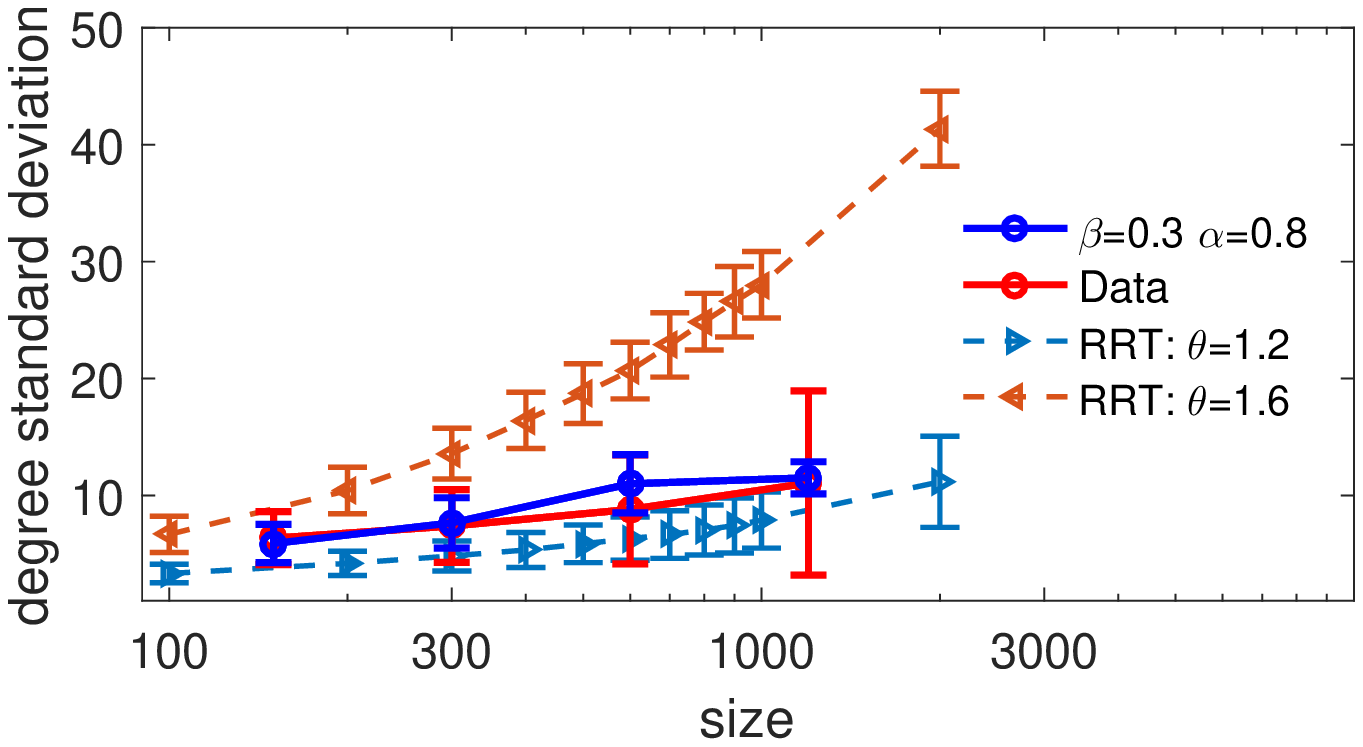}
			\label{fig:SeirTopoAna1b}
		}	
		\subfigure[]{
			\includegraphics[scale=.4]{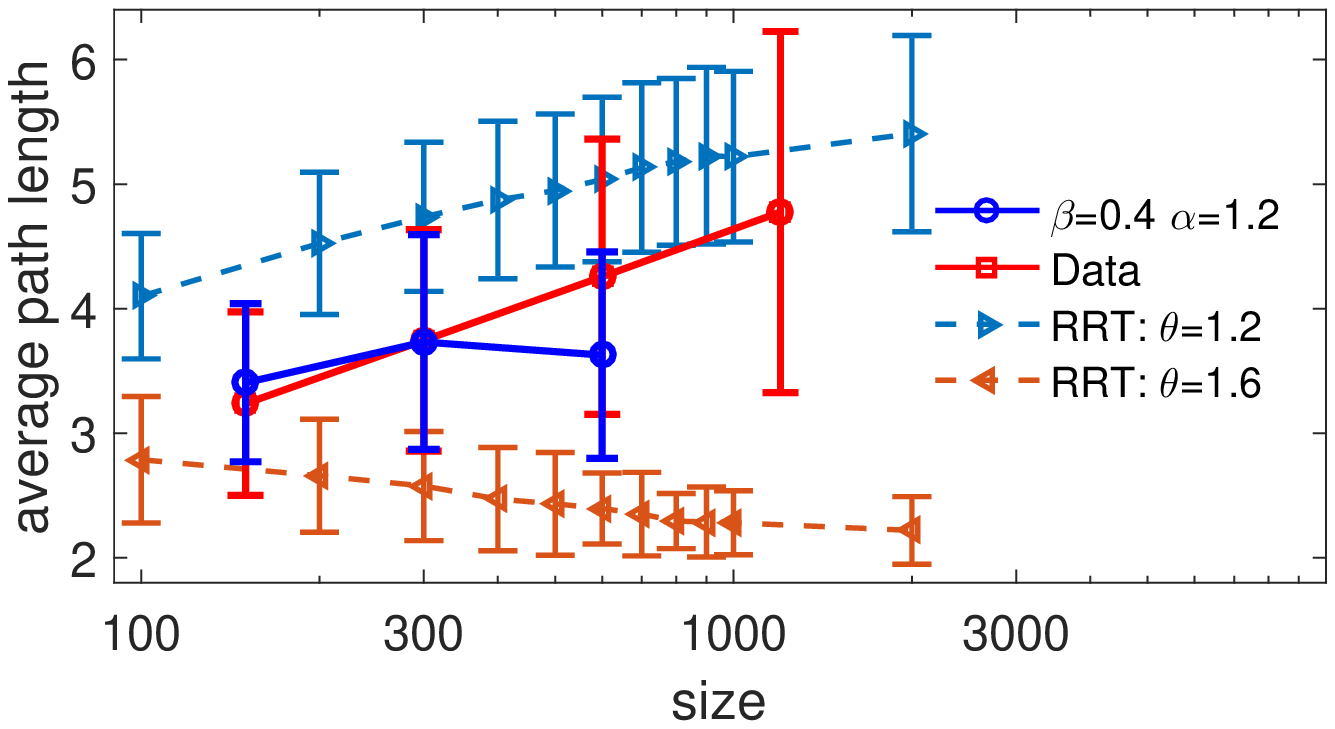}
			\label{fig:SeirTopoAna1c}		
		}
		\subfigure[]{
			\includegraphics[scale=.4]{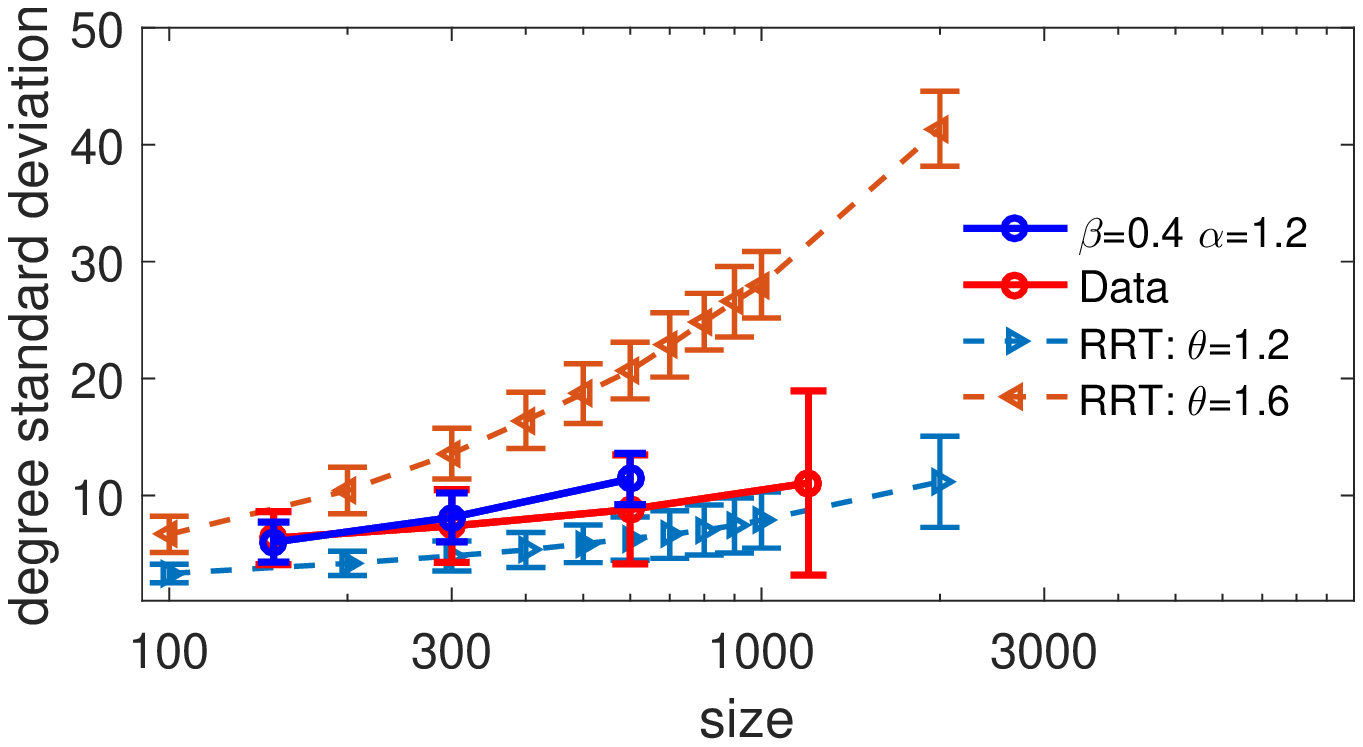}
			\label{fig:SeirTopoAna1d}
		}
		\subfigure[]{
			\includegraphics[scale=.4]{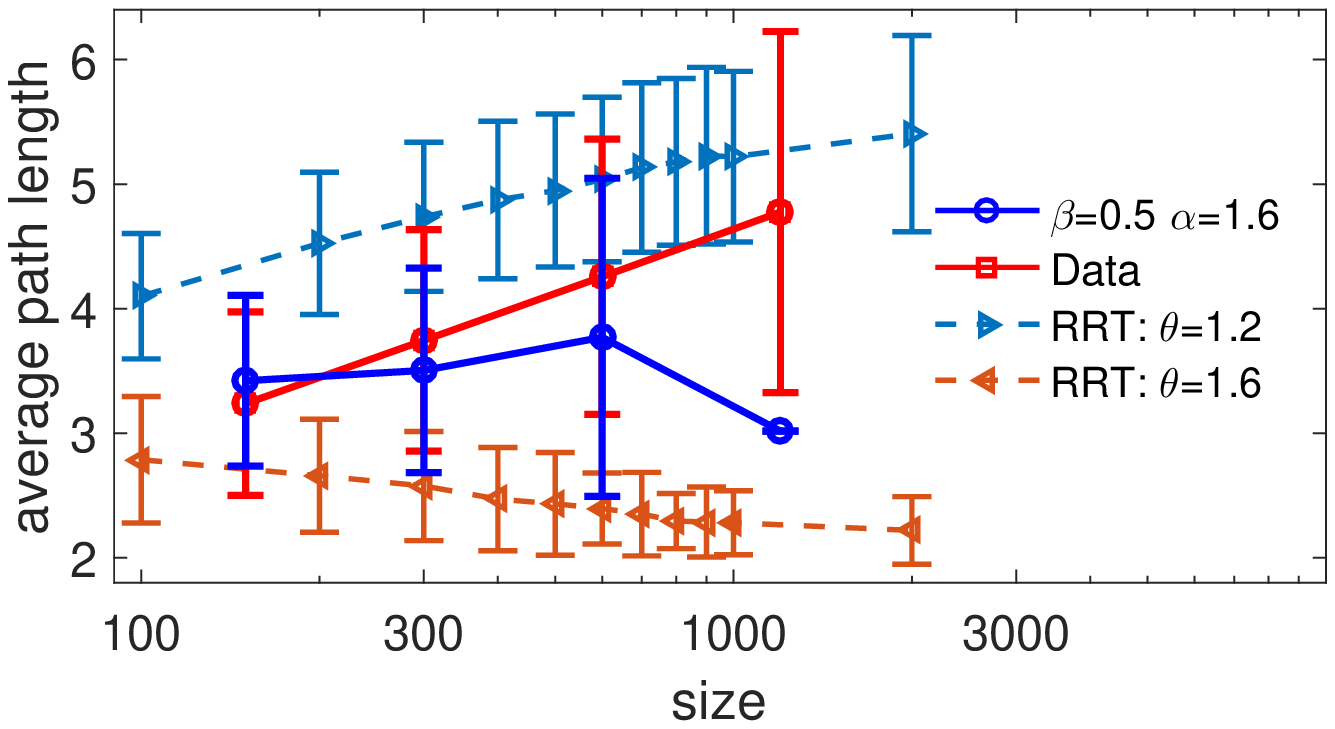}
			\label{fig:SeirTopoAna1e}		
		}
		\subfigure[]{
			\includegraphics[scale=.4]{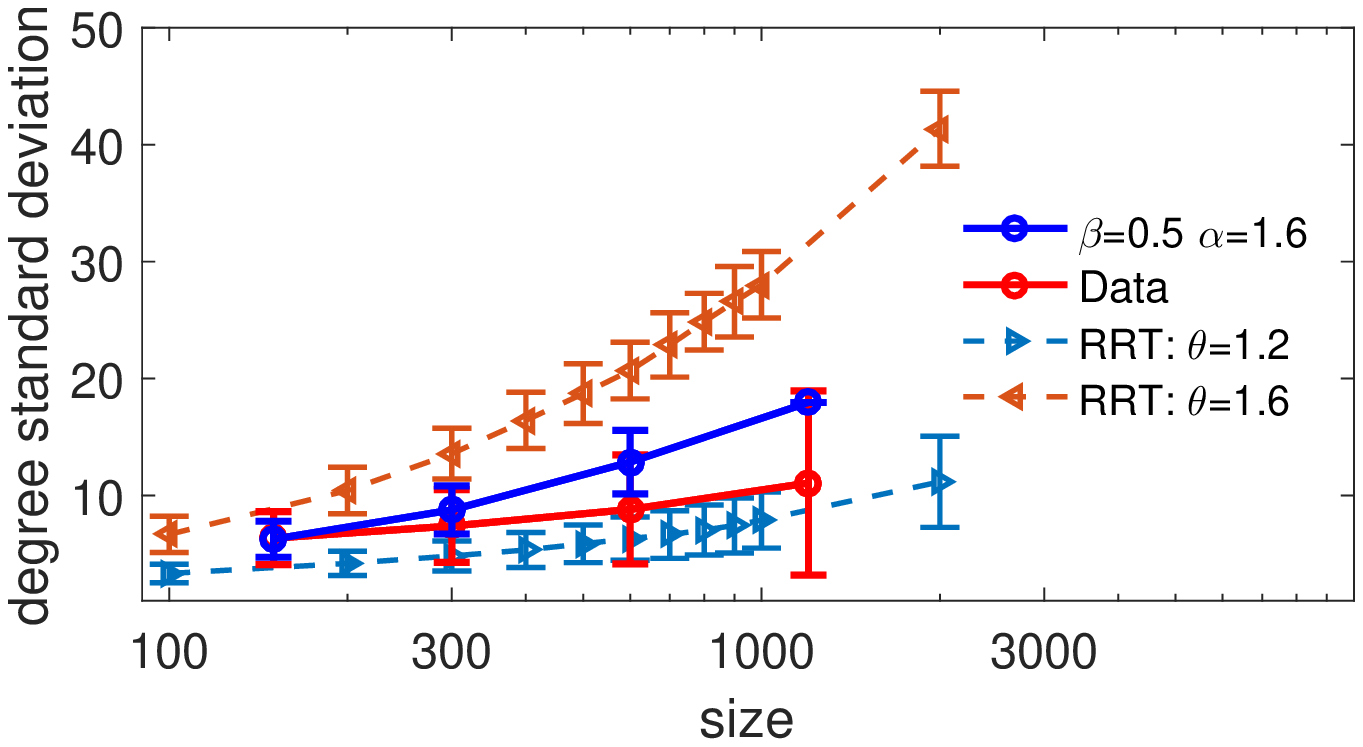}
			\label{fig:SeirTopoAna1f}
		}	
		\caption{\csentence{The average path length and degree standard deviation of the cascade trees in WeChat, of the RRT structural model and of the cascade trees generated by the SVFR model.} We consider the SVFR model with the three sets of parameters $\beta$ and $\alpha$ that could well capture the WeChat cascade size distribution. The underlying networks of the SVFR model are scale-free with size $N=10^5$. Given the parameter $\beta$ and $\alpha$, we perform $100$ realisations of the SVFR model on each of the $100$ independently generated underlying networks leading to $10^4$ cascade trees. These cascade trees generated by SVFR are grouped according to their sizes: [100,200), [200,400), [400,800) and [800,1600]. The average and standard deviation of the two key properties are deived for each group and plotted as a function of the medium size of the group. When $\beta=0.4$ and $\alpha=1.2$, the cascade trees generated by SVFR model are all smaller than 800 in size. Given the parameter $\theta$ and tree size, we carry out $10^3$ iterations of generating the cascade trees using RRT model and obtain the average and standard deviation (error bar) of these two properties. }
		\label{fig:SeirTopoAna1}
	\end{figure}
	
	Figure \ref{fig:SeirTopoAna1} shows that the cascade trees generated by the SVFR model with $\beta=0.3$ and $\alpha=0.8$ well approximate the cascade trees in WeChat with respect to their average path length and the degree variance/standard deviation. The cascade trees generated by the SVFR, the same as the WeChat cascade trees, are also well bounded by the RRT models with $\theta=1.2$ and $\theta=1.6$ and closer to RRT models with $\theta=1.2$, verifying the consistency of the RRT and SVFR models. 
	
	As mentioned before, it would be interesting to explore even larger underlying network sizes, which could lead to larger cascade trees thus improve the SVFR model validation with respect to capturing features of cascade trees with a broader range of sizes.
	
	Our SVFR model could well explain the cascade size distribution including the power-law decay exponent, the average path and the degree variance of the cascade trees in WeChat and suggests that a user with a large number of friends may have a lower probability to view the message shared by a friend.
	
	\section{Conclusion}
	\label{sec:conclusion}
	
	The cascade trees that describe the information spread trajectories in social networks have been widely studied. In this work, we rely on the data extracted from the WeChat social network as a test bed to further advance the information diffusion analysis methods from two aspects. 
	
	Firstly, we propose to model the cascade tree topology by random recursive trees RRTs. The RRT model could well reproduce or explain two fundamental properties of the cascade trees in the WeChat network, i.e. the average path length and the degree variance in relation to the tree size. The identified single parameter $\theta$ in the RRT model, allows us, for the first time to quantify how deep (viral like spread) or shallow (broadcast type spread) a class of cascade trees are. Hence, we could compare or classify different online networks regarding to that the information spread on each network is more broadcast or viral like. The RRT model also unravels some interesting phenomena in the cascade-tree growth, like the emergence of hubs. 
	
	Secondly, we introduced the SVFR stochastic model to capture the information diffusion process on a network. The model encodes three types of user reactions to a message they receive: ignore, view or forward the message, and was shown to capture and explain three main properties of the WeChat cascade trees: the average path length, the degree variance and the tree size distribution. Our model calibration suggests that a WeChat user with a large number of friends tends to have a low probability to view a message shared by his/her friends. This finding can be supported by the cognitive and biological constraints of users as predicated by Dunbar's theory \cite{DUNBAR,Perra2011}.
	
	The WeChat dataset served as excellent test bed enabling the above mentioned contributions due to the rich user actions it captures and related to the way how users react to the message forwarded to them. We believe, however, that our contributions can serve as a starting point to systematically explore the structure and dynamics of information diffusion in general social networks, not limited to WeChat. The proposed SVFR stochastic model can be applied to other online social networks as well to explore e.g. whether other types heterogeneity may exist. For example, the view or forward probability of a content may depend on the content. Another promising future research direction is to explore the time delay in the information diffusion model in order to explain e.g. how fast a message could reach a certain number of users.

	\begin{backmatter}
		
		\section*{Competing interests}
		The authors declare that they have no competing interests.
		
		\section*{Author's contributions}
		Conceived and designed the experiment: LL, BQ, BC, HW. Performed the experiment: LL. Analyzed the data: LL, BQ, AH and HW. Wrote the paper: LL and HW. All authors read and approved the final manuscript.		
		
		\section*{Acknowledgements}
		The authors would like to thank Linnan He (The School of Communication and Design, Sun yat-sen University) and Yichong Bai (Fibonacci Consulting Co. Ltd.) for providing the WeChat dataset. We also wish to thank the National Natural Science Foundation of China under Grant No. 71673292, 61503402, 61403402, 61374185, 71373282. 
		
		\bibliographystyle{bmc-mathphys} 
		\bibliography{ICMEJPDS}      
		
		
		
		

	\end{backmatter}
\end{document}